\documentclass[10pt, a4paper, twocolumn, aps, pra, showpacs, longbibliography, nofootinbib,superscriptaddress]{revtex4-1}
\pdfoutput=1
\usepackage[utf8x]{inputenc}
\usepackage{ucs}
\usepackage{amsmath}
\usepackage{amsfonts}
\usepackage{amssymb}
\usepackage{makeidx}
\usepackage{cellspace,booktabs}
\usepackage{natbib}
\usepackage{lipsum}
\usepackage{bm}
\usepackage{bbm}
\usepackage{relsize}
\usepackage{float}
\usepackage{wrapfig}
\usepackage{units}
\usepackage{multirow}
\usepackage{braket}
\usepackage{mathtools}

\usepackage{hyperref}
\hypersetup{colorlinks = true, linkcolor=magenta,citecolor=blue, urlcolor=magenta, bookmarksnumbered = true}

\usepackage{color}
\definecolor{light-gray}{gray}{0.55}

\usepackage{microtype}

\usepackage{graphicx}

\usepackage{xcolor}
\usepackage[normalem]{ulem}

\begin{document}

\begin{abstract}
Bilayer graphene is a nanomaterial that allows for well-defined, separated quantum states to be defined by electrostatic gating and, therefore, provides an attractive platform to construct tunable quantum dots. When a magnetic field perpendicular to the graphene layers is applied, the graphene valley degeneracy is lifted, and splitting of the energy levels of the dot is observed. Although bilayer graphene quantum dots have been recently realized in experiments, it is critically important to devise robust methods that can identify the observed quantum states from accessible measurement data. Here, we develop an efficient algorithm for extracting the model parameters needed to characterize the states of a bilayer graphene quantum dot. Specifically, we put forward a Hamiltonian-guided random search method and demonstrate robust identification of quantum states on both simulated and experimental data.
\end{abstract}

\date{\today}
\author{Jozef Bucko}
\email{jozef.bucko@uzh.ch}
\affiliation{Institute for Theoretical Physics, ETH Zurich, CH-8093, Switzerland}
\affiliation{Institute for Computational Science, University of Zurich, Winterthurerstrasse 190, 8057 Zurich, Switzerland}
\author{Frank Schäfer}
\address{Department of Physics, University of Basel, Klingelbergstrasse 82, CH-4056 Basel, Switzerland}
\author{František Herman}
\affiliation{Department of Experimental Physics, Comenius University, Mlynská Dolina F2, 842 48 Bratislava, Slovakia}
\affiliation{Solid State Physics Laboratory, ETH Zurich, CH-8093 Zurich, Switzerland}
\author{Rebekka Garreis}
\affiliation{Solid State Physics Laboratory, ETH Zurich, CH-8093 Zurich, Switzerland}
\author{Chuyao Tong}
\affiliation{Solid State Physics Laboratory, ETH Zurich, CH-8093 Zurich, Switzerland}
\affiliation{Institute for Theoretical Physics, ETH Zurich, CH-8093, Switzerland}
\author{Annika Kurzmann}
\author{Thomas Ihn}
\affiliation{Solid State Physics Laboratory, ETH Zurich, CH-8093 Zurich, Switzerland}
\author{Eliska Greplova}
\affiliation{Kavli Institute of Nanoscience, Delft University of Technology, Delft, the Netherlands}
\affiliation{Institute for Theoretical Physics, ETH Zurich, CH-8093, Switzerland}

\title{Automated reconstruction of bound states in bilayer graphene quantum dots}

\maketitle

\section{Introduction}

Atomically thin nanomaterials provide an exciting platform for quantum technologies~\cite{novoselov20162d, liu2016van, liu20192d, kennes2021moire}. Bilayer graphene has specifically drawn great attention~\cite{ohta2006controlling, mccann2013electronic, Bucko_2021, cao2018unconventional,PhysRevLett.123.096802} and it was shown that bilayer graphene is a promising host for gate-defined quantum dots~\cite{eich_2018}. Quantum dots are one of the prime candidates for scalable and highly controllable quantum devices~\cite{hill2015surface, watson2018programmable, Loss_1998, Trauzettel_2007}. Quantum dot technology has benefited from the advances in material science and has led specifically to the fabrication of high-quality bilayer graphene devices, which provide a range of benefits for quantum dot applications~\cite{eich2018coupled, kurzmann2019excited, Greplova_2020, banszerus2020dispersive}. 

The two dominant sources of decoherence for spin qubits in quantum dots are spin-orbit coupling and hyperfine coupling of nuclear and electronic spins, both of which are expected to be largely suppressed in graphene~\cite{lyon2017probing, sichau2019resonance, Min_2006, Hernando_2006}. Moreover, it is possible to control the size of the gap in bilayer graphene via a vertical electric field~\cite{Ohta_2006, rickhaus2019gap}, which has been successfully used for  charge carrier confinement.

\begin{figure}
\centering
\includegraphics[scale=1]{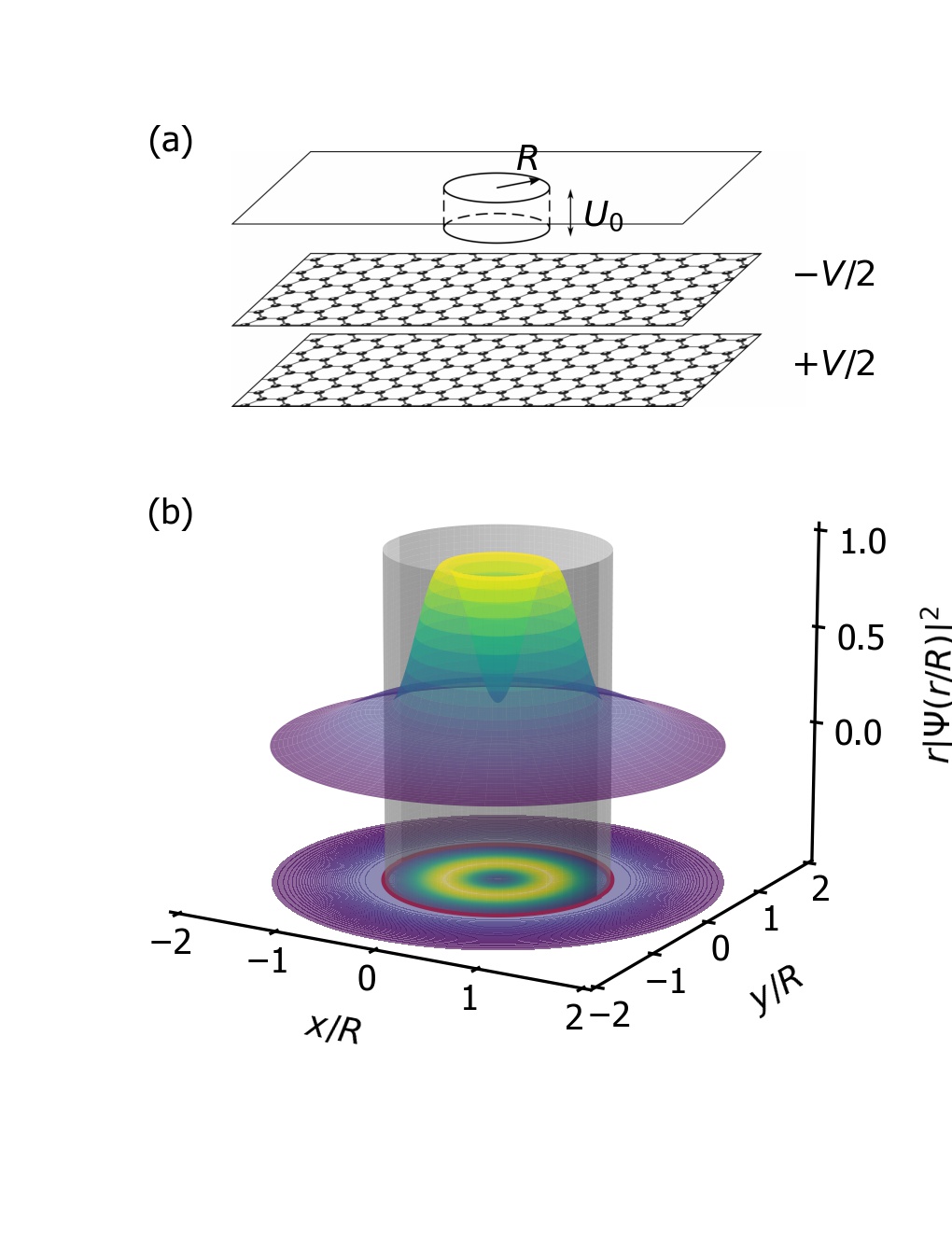}
\caption{(a) Illustration of confining potential $U$ [$U_0=U(R)$] and gapping potential $V$ for a quantum dot realized in a bilayer graphene;
(b) Visualization of the bound state wavefunction for $m=0,\tau = 1, U=60\, \unit{meV}, V = 50\, \unit{meV}$ (top). The projection of the wavefunction is shown in the bottom part of the plot with the gray cylinder indicating the edge of the quantum dot. The red circle indicates the edge of the quantum dot in the projected plane.}
\label{fig:intro}
\end{figure}

An overarching goal of quantum dot engineering is the design of qubits that can be used for quantum information processing. While spin qubits are straightforward to engineer in silicon and have achieved high-fidelity quantum operations~\cite{zwanenburg2013silicon, veldhorst2014addressable, yoneda2018quantum}, they do not yet meet the demands of scalable quantum devices. Modern approaches to the design of semiconductor qubits therefore utilize the theoretical knowledge of the internal state structure of quantum dots in order to define qubits with longer coherence times~\cite{landig2018coherent, cerfontaine2020closed, kratochwil2020realization}. While bilayer graphene has excellent electronic properties, the theoretical understanding of the band structure of multilayer graphene in the presence of electric and magnetic fields is not yet fully developed. However, such an understanding is critically needed to design and define qubits with optimal coherence properties. Currently, simple models of graphene quantum dots already capture many physical aspects of these devices~\cite{recher_2009,mccan2006,zarenia2013},
but the question of how to reconcile these predictions with experimental observations remains a challenge.

In this work, we address the challenge of connecting transport measurements in bilayer graphene quantum dots (BGQD)~\cite{eich_2018} to an underlying theoretical description of quantum states that a spin can occupy inside a bilayer graphene quantum dot. First, we introduce a computational framework for the interpretation of transport measurements in terms of a detailed bilayer graphene quantum dot bound state characterization. We then apply this framework to both simulated data and experimental measurements of the bilayer graphene quantum dots. Our algorithm, Hamiltonian-driven random search (HRS), leverages a combination of adjoint methods and global optimization to navigate the complex structure of the multiparameter optimization landscape with the ultimate goal of identifying an optimal model describing the measured data. Specifically, our HRS framework combines the Hellmann-Feymann theorem applied on a candidate Hamiltonian and controlled random search, ultimately taking advantage of exact model-based gradients to confine a well-defined region for the random search, thus leading to higher accuracy.

This manuscript is organized as follows: In Section~\ref{sec:bi} we review a low-energy theory for the description of bound states in a bilayer graphene quantum dot. 
In Section~\ref{sec:HRS} we present the Hamiltonian-guided random search algorithm. In Section~\ref{sec:results} we demonstrate HRS performance on numerically generated data based on the low-energy theory and benchmark it against standard non-Hamiltonian driven random search methods. Additionally, we apply HRS on experimentally measured data to obtain a complete wave-function characterization of the measured states. We discuss assumptions and limitations of our work in Section~\ref{sec:disc}. Finally, we present a conclusion, a discussion, and an outlook on future applications of our method in experimental settings in Section~\ref{sec:conclusion}.

\section{Bilayer Graphene Quantum Dots}
\label{sec:bi}

In what follows, we introduce a simple theory description which is able to predict the structure of the single-particle bound states in bilayer graphene together with the structure of the respective energy levels. Our theory description is based on the energetically most favorable \cite{rozhkov_2016} (and therefore experimentally most relevant) structure of the bilayer graphene, where the individual graphene layers are in the AB or Bernal stacking geometry. The model we consider, first introduced in \cite{recher_2009}, also allows us to include the effect of the external voltages applied through electrostatic gates as well as the effect of the external magnetic field. Due to its simplicity this model does not include effects related to the spin of the particle and also assumes perfect cylindrical symmetry of a quantum dot. Here, our goal is not to find the most physically exhaustive description of the bilayer graphene quantum dots, but rather, to build a compact effective model that can be scalably fitted to experimental data and decisively determine the underlying bound state structure in the quantum dot.

In bilayer graphene quantum dots, the energy of bound states is defined via an interplay of two dominant energy scales. First, the energy is dominated by the inter-layer hopping strength, which has typical values of $\sim 400$~meV~\cite{mccann2013electronic}. Second, a gapping  potential $V$, which results from an interlayer electrostatic potential asymmetry, opens an energy gap in the spectrum. Recent experiments~\cite{eich_2018} determined a typical value of this energy gap of about $\sim 60$~meV. For (bilayer) graphene, the Fermi energy is on the order of $\sim 7.5 - 9$ eV. 
Notably, the low-energy bound states are expected to have energies that are smaller than the energy gap and thus also much smaller than the Fermi energy.
Therefore, we expect the bound states to have momenta close to the Dirac points of graphene such that we can linearize the dispersion relation around these points to obtain an effective description of the system in this low-energy regime.

A semi-analytically solvable theoretical description of bilayer graphene quantum dots was put forward in Ref.~\cite{recher_2009}. The combination of the above-mentioned physically motivated linearization of the dispersion relation, a tight-binding approximation, and specific restrictions on the symmetry of the quantum dot yields the following Hamiltonian in first quantization:
\begin{equation}
\label{eq:H0b_1}
\mathcal{H}=
\begin{pmatrix}
U(r) + \frac{\tau V}{2} & p_x + i p_y & t_{\perp} & 0 \\
p_x - i p_y & U(r) + \frac{\tau V }{2} & 0 & 0 \\
t_{\perp} & 0 & U(r) -\frac{ \tau V}{2} & p_x - i p_y \\
0 & 0 & p_x + i p_y & U(r) -\frac{ \tau V}{2}
\end{pmatrix},
\end{equation}
where $U,V$ is the confining and gapping voltage (see Fig.~\ref{fig:intro}), respectively, $r = \sqrt{x^2 + y^2}$ is the radial coordinate for the potential $U$, $\tau$ is the valley quantum number, $t_\perp$ is the energy associated with inter-layer hopping (we use $t_\perp = 400$~meV), and $p_i$ are the momentum operator components. 

\begin{figure}
\centering
\includegraphics[scale = 1.]{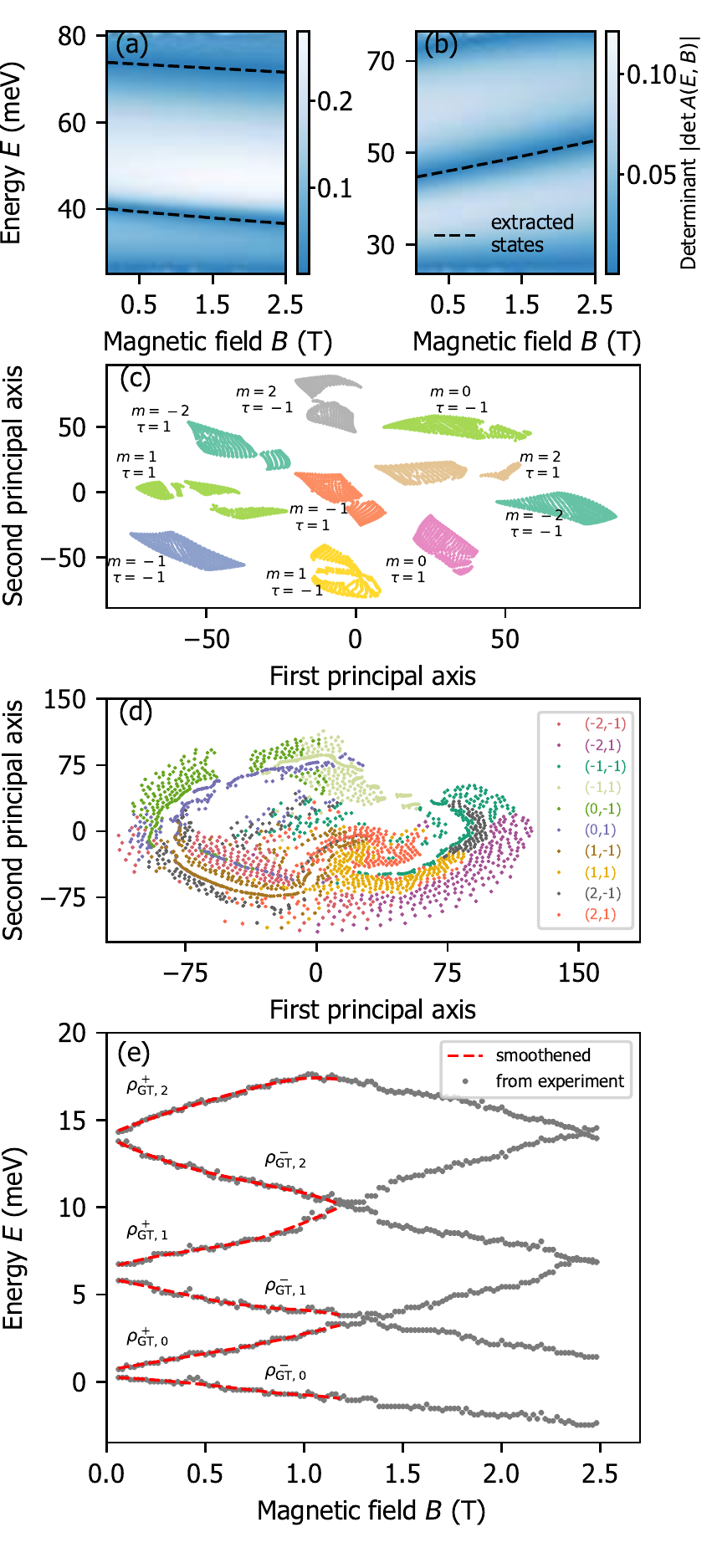}
\caption{Difference between information content of the theoretical model and the experimental measurement. (a), (b) values of the determinant, det $|A(E, B)|$ as a function of energy, $E$, and magnetic field, $B$, with $m=0,\tau = -1, U=53\, \unit{meV}, V = 54\, \unit{meV}$ for panel (a) and $m=2,\tau = 1, U=50\, \unit{meV}, V = 51\, \unit{meV}$ for panel (b). Black dashed lines represent extracted energy states and correspond to the experimentally accessible information. Panel (c) shows the t-SNE clustering applied on the set of determinant maps like those shown in (a) and (b). Each cluster corresponds to a distinct pair of angular momentum, $m$, and valley, $\tau$, quantum numbers. Panel (d) shows the result of t-SNE clustering on the eigenenergy lines as highlighted by the dashed lines in (a), (b). Panel (e) displays the pre-processed experimental data for one of measured dot systems (grey) and the resulting smoothened data used within the Hamiltonian parameter reconstruction (red).
}
\label{fig:TSNE}
\end{figure}

Under the assumption of a perfectly circularly fabricated quantum dot, we can enforce cylindrical symmetry of the wave spinor that yields the factorization of the wave function into radial and orbital parts
\begin{equation}
\label{eq:transform1}
\Psi(r,\varphi) = 
\frac{e^{ i m \varphi}}{\sqrt{r}}
\begin{pmatrix}
1 & 0 & 0 & 0 \\
0& e^{- i \varphi} & 0 & 0 \\
0& 0 & 1 & 0 \\
0 & 0 & 0 & e^{ i \varphi}
\end{pmatrix}
\Psi_1(r),
\end{equation} 
where $\Psi_1(r)$ is the radial contribution of the spinor and $m$ is an angular momentum quantum number.

Reference~\cite{recher_2009} further assumes a continuity of the wave function on the boundary of the quantum dot (assumed to be at $R=20$~nm from the dot centre throughout this work), which reduces the eigenenergy problem to a set of linear equations for the components of the spinor $\Psi_1(r)$. The solution of the system of linear equations can be formulated as a zero-determinant condition. We describe details of this solution in Appendix~\ref{app:model_analytical}. 

In Fig.~\ref{fig:TSNE}(a), (b) we show examples of the determinant values for the eigenvalue problem associated with the Hamiltonian in Eq.~\eqref{eq:H0b_1} as a function of energy, $E$ and  magnetic field, $B$. The darkest blue lines correspond to the points where the determinant values are zero and therefore correspond to energy eigenvalues of the problem. These lines thus also determine the theoretical prediction of the energies that are accessible via transport measurements of the dot in the perpendicular magnetic field~$B$. The exact values of the eigenenergies are depicted as black dashed lines.

By reformulation of the model of Eq.~\eqref{eq:H0b_1} into the determinant condition shown in Fig.~\ref{fig:TSNE}(a), (b) we unveil a certain amount of information about the physics of the system. Specifically, using an existing clustering technique, the t-Distributed Stochastic Neighbor Embedding (t-SNE)~\cite{maaten_2008_tsne_main}, on a set of determinants generated for various configurations of discrete quantum numbers, $m$ and $\tau$, and various potential values, $U$ and $V$, we find that all possible combinations of the quantum numbers are clearly distinguished [see Fig.~\ref{fig:TSNE} (c)].

Applying the same clustering on the subset of the data that is accessible experimentally [the dashed eigenenergy lines in Fig.~\ref{fig:TSNE}(a), (b)], we obtain a two-dimensional embedding shown in~Fig.~\ref{fig:TSNE}(d). We observe that distinct clusters of possible quantum number pairs are no longer identifiable. In Appendix~\ref{app:unsupervised} we describe in more detail the low-dimensional embedding of the model output, and in Appendix~\ref{app:nn_det} we provide a further discussion and other methods for extracting the wave function and Hamiltonian parameters from the gradient profile of the determinant maps.

In Fig.~\ref{fig:TSNE}(e), we show an example of experimental data used in our study as resulting from the transport measurement of the sample presented in Ref. \cite{eich_2018}. In the experiment, the quantum dot is defined and tuned through electrostatic gates deposited on top of a stack of hBN-bilayer graphene-hBN. The energy levels of the quantum dot are extracted from peaks in the conductance through the area of the device where the quantum dot is defined. An essential difference to the data presented in \cite{eich_2018} is an additional post-processing step we performed. Specifically, we average over the spin degree of freedom in the measured data. The reason for this is that the simplified model of Eq.~\eqref{eq:H0b_1} depends only on the valley degree of freedom and not on the spin. This averaging results in two degenerate states (at zero magnetic field) per energy corresponding to two valley states $\tau=\pm 1$. We explain our treatment of the experimental data in detail in Appendix \ref{app:exp_data}.

Up to this point, we have demonstrated that even if we radically simplify the description of our system to only capture its main characteristics, the reconstruction of this description from accessible experimental data only is potentially demanding. In the following, we will introduce a hybrid optimization method that straightforwardly allows for the inference of the Hamiltonian parameters and wave function based on transport measurement data, even in the case of extremely challenging optimization landscapes.

Throughout the text, we abbreviate the full set of Hamiltonian parameters as $\rho = (m,\tau,U,V)$. We further distinguish the Hamiltonian parameters resulting from the optimization process $\rho_{{\rm opt}} = (m_{{\rm opt}},\tau_{{\rm opt}},U_{{\rm opt}},V_{{\rm opt}})$ and ground truth, sought-after parameters $\rho_{{\rm GT}} = (m_{{\rm GT}},\tau_{{\rm GT}},U_{{\rm GT}},V_{{\rm GT}})$. More specifically, $\rho_{\rm GT}$ can denote either ground truth values used in the theoretical model \eqref{eq:H0b_1} or the underlying (assumed) values of experimental data we aim to approach by the $\rho_{\rm opt}$ set. The parameters $\rho_{\rm GT}$ are known in the case of simulated data and they are unknown in the case of experimentally measured data.  Moreover, where relevant we distinguish increasing (decreasing) energy states by superscript $s=+(-)$. When talking about experimental data, it will be necessary to label multiple pairs of approaching states as $B\rightarrow 0$~T [see, e.g., Fig.~\ref{fig:TSNE}(e)]. In such a case, we add an integer index to the above notation. For example, $\rho_{{\rm opt},0}^+ = (m_{{\rm opt},0}^+,\tau_{{\rm opt},0}^+,U_{{\rm opt},0}^+,V_{{\rm opt},0}^+)$ denotes optimal (found) Hamiltonian parameters for an increasing ($+$) branch of lowest energy state pair ($i=0$).

\begin{figure}
    \centering
    \includegraphics[width=\linewidth]{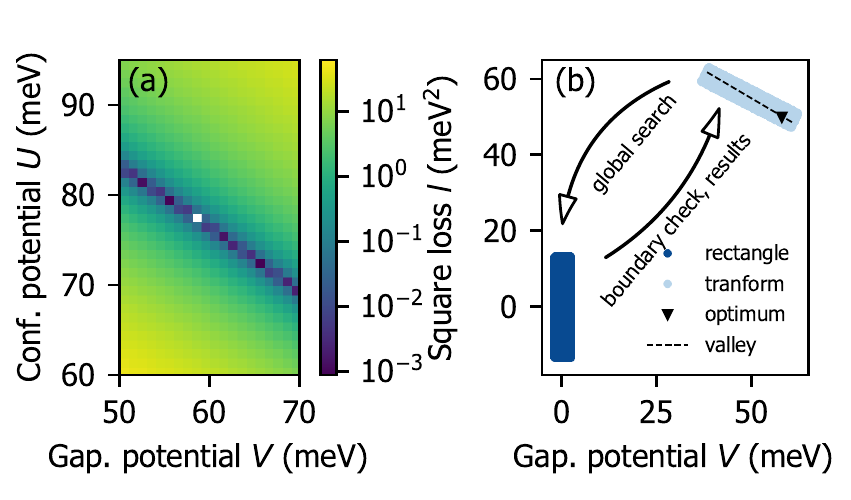}
    \caption{(a) 
    The $U-V$ optimization landscape with model-based target state $m_{\rm GT}~=~0,U_{\rm GT}=78\,\unit{ meV}, V_{\rm GT} = 58\, \unit{meV},\tau_{\rm GT} = 1$. The narrow valley defining the search domain of the global optimization algorithm is visible.
    (b) Search domain transformation scheme for the global  optimization subroutine. Once the confined domain for the global optimization is found (light blue), this domain is aligned with the coordinate along $U$ and $V$ (dark blue), and the global optimization algorithm (CRS-LM) is applied. Afterwards, the domain is transformed back. }
    \label{fig:landscape_transform}
\end{figure}

\section{Hamiltonian-guided optimization}
\label{sec:HRS}

We aim at solving the following optimization task: We want to infer the four unknown Hamiltonian parameters such that the zero-determinant dashed line in Fig.~\ref{fig:TSNE}(a), (b) will fit the measured energies shown in Fig.~\ref{fig:TSNE} (e). However,  the parameter landscape exhibits a series of neighboring shallow local minima with very small differences in energy distributed in a confined domain of the parameter space [see Fig.~\ref{fig:landscape_transform}(a)]. This type of landscape immediately excludes any local, gradient-based optimization method. The series of shallow adjacent minima is extremely adversarial for systematic exploration -- the gradient methods get systematically stuck in one of the local minima. At the same time, the precision of gradient-free methods suffers from the size of the multi-dimensional parameter space.

Here we want to note the advantageous interplay between the structure of our model and complexity of optimization landscape. The four-parameter Hamiltonian in Eq.~\eqref{eq:H0b_1} can be written as $4\times4$ matrix and provides a controllable toy model that allows us to benchmark our optimization method without too high computational overhead. At the same time, this comparatively simple model exhibits an optimization landscape complexity that is challenging for gradient methods.

We begin by dividing the task of inference of the underlying system parameters into two parts: (i) the determination of the continuous confining and gapping potentials $U$ and $V$, and (ii) the determination of the discrete quantum numbers $m$ and $\tau$.

An example of the optimization landscape for the regression of the potentials $U, V$ is shown in Fig.~\ref{fig:landscape_transform}(a). One can observe a long, shallow valley of local minima with the white square representing the global optimum at the ground-truth values $U_{\rm GT}, V_{\rm GT}$. This shallow minimum structure exhibiting noisy features prevents gradient-based optimization methods from working effectively. Thus, a global optimization method is needed. While global methods work well in landscapes without a clear dominant gradient profile, they can be computationally expensive and often do not perform optimally over large connected domains within the full optimization space~\cite{LSGO_benchmark_2013}. 

We formulate the combination of a global and a local method that retains the advantages of the global search but uses the local optimization subroutine to alleviate the computational cost of the global method. Specifically, we use our knowledge of the Hamiltonian in Eq.~\eqref{eq:H0b_1} to dramatically restrict the domain of the global method into the shallow minimum and there enhance its precision and efficiency.

Generally, to identify the valley of local minima in the energy landscape (in our case parametrized by gapping and confining potential), we employ the mean squared error loss defined as
\begin{equation}
    \label{eq:square_loss}l_{m,\tau}(U,V) = \frac{1}{B_{\rm max}}\int_{0}^{B_{\rm max}}\left( E^{m,\tau} - E_{\rm GT}^{m,\tau}\right)^2 {\rm d}B. 
\end{equation}
Here, $E^{m,\tau}$ is the energy as a function of the applied magnetic field, $B$, for fixed $m, \tau$ and generic potential value $U,V$.  $E_{\rm GT}^{m,\tau}$ denotes the desired energy we wish to optimize for: it can either be extracted from the simulated data using the determinant condition or directly experimentally measured. The loss $l_{m,\tau}(U,V)$ measures the difference of the optimized and ground-truth energy across the range of magnetic field.

The computation of derivatives of the loss function in Eq.~\eqref{eq:square_loss} with respect to the parameters $U, V$ is required in order to implement a local, gradient-based subroutine that identifies the region of shallow minima. Such derivatives could be in principle approximated using numerical differentiation, which, however, suffers from floating-point and truncation errors. We can avoid numerical differentiation altogether through the application of adjoint methods~\cite{cao_2003_adjoint_method} on the problem Hamiltonian.

Specifically, we derive the gradients with respect to $U$ and $V$ analytically using the Hellman-Feynman theorem \cite{feynman_1939_HF}. For $Q\in\{U,V\}$, we obtain
\begin{align}
    \label{eq:H-F_for_loss}
    &\frac{\partial l_{m,\tau}(U,V)}{\partial Q} = \frac{\partial l_{m,\tau}(U,V)}{\partial E^{m,\tau}}\frac{\partial E^{m,\tau}}{\partial Q}=\nonumber\\ &\frac{1}{B_{\rm max}}\int_{0}^{B_{\rm max}} 2(E^{m,\tau} - E_{\rm GT}^{m,\tau}) \Braket{\Psi | \frac{\partial H}{\partial Q} | \Psi}{\rm d}B,
\end{align}
where $\Psi$ is a wave spinor as defined in Eq.~\eqref{eq:transform1}. 

Once we identify the boundaries, using the gradient-based method, of the shallow minima shown in Fig.~\ref{fig:landscape_transform} (a), we can initialize the global optimizer on a much better confined domain.
Based on state-of-the-art benchmarks of global optimization algorithms~\cite{Arnoud2019BenchmarkingGO} and the profile of our landscape, we choose the controlled random search with local mutations as our global optimizer (CRS-LM)~\cite{kaelo_ali_2006_crs-lm}. This optimization method comprises of the following steps: Starting from the initialization of a set of random points, a simplex is constructed from a subset of the random points with the associated smallest loss function values, see Eq.~(\ref{eq:square_loss}). New trial points are then generated upon reflections with respect to the simplex. The local mutation implies that unsuitable points are not discarded but modified via a specific mutation condition. We provide a detailed description of the algorithm in Appendix~\ref{app:crs}. In our work, we use the implementation within the NLopt package \cite{johnson_nlopt}.

The optimization domain is most natural to work with when aligned with coordinate axes. Therefore, we transform the original domain [light blue in Fig.~\ref{fig:landscape_transform}(b)] to be centered at zero and aligned with the $U,V$ axes [dark blue in Fig.~\ref{fig:landscape_transform}(b)] during the optimization. Throughout the optimization, we apply the corresponding inverse transformation to retrieve the original domain in order to perform boundary checks and to store the results (see Appendix~\ref{app:hrs} and link in Ref.~\cite{BGQD_code}).

The HRS algorithm thus represents a computationally efficient modification of the global CRS method for landscapes manifesting shallow minima profile. This modification is possible due to our knowledge of the candidate Hamiltonian model. This knowledge allows for the physics-guided implementation of the gradient-free global method. Ultimately, as we show using our numerical results in the next section, this method materially reduces the average error of the reconstructed parameters.

In the algorithm description so far, we optimized the continuous parameters $U, V$ while assuming that the discrete parameters $m, \tau$ were arbitrary but fixed. Fortunately, the discrete parameters present only a small set and thus little added optimization complexity. We proceed by determining all relevant candidate pairs $(m, \tau)$ and re-run HRS as described above for all these pairs. Comparing the gradient of measured data and theoretical simulation of single-particle energy lines, we conclude it is sufficient to explore momenta $m=\{-3,-2,...,3\}$ for both positive and negative valley number $\tau$.

Specifically, to fit the discrete parameters, we need to minimize the total cost function for all states we are fitting. This is done in order to find the unique combination $(m,\tau)$ for each state being fitted that leads to the smallest mean squared error, i.e.,
\begin{align}
\label{eq:big_loss}(m_{{\rm opt},j}^\pm, \tau_{{\rm opt},j}^\pm)
    =\underset{\substack{m \in \mathbb{Z}\setminus \{0\} \\ \tau\in\{\pm 1\}}}{\rm argmin} \sum_{i=1}^N \sum_{s\in\{\pm\}} l_{m_i^s,\tau_i^s} \left(U^{s}_{{\rm opt},i},V^{s}_{{\rm opt},i}\right).
\end{align}
Here, $j \in \{1,\dots,N\}$ and $l_{m_i^s,\tau_i^s} \left(U^{s}_{{\rm opt},i},V^{s}_{{\rm opt},i}\right)$ denotes the loss defined in Eq.~\eqref{eq:square_loss} evaluated for the continuous parameters $U^{s}_{{\rm opt},i},V^{s}_{{\rm opt},i}$, which results from the HRS optimization in the $U-V$ plane given a discrete pair $(m_i^s,\tau_i^s)$. The inner sum is performed over $N$ different  $s=\pm$, where we sum over the valley-splitted pair of states. Due to time reversal symmetry \cite{recher_2009}, the quantum number for each such pair fulfill $m_i^+ = -m_i^-$ and $\tau_i^+  = -\tau_i^-$. Note that this symmetry constraint restricts the total number of possible combinations of discrete quantum numbers. The outer sum then corresponds the summation over all $N$ pairs of energy lines (i.e., $2N$ lines in total).

\section{Results}
\label{sec:results}

\begin{figure}
    \centering
    \includegraphics{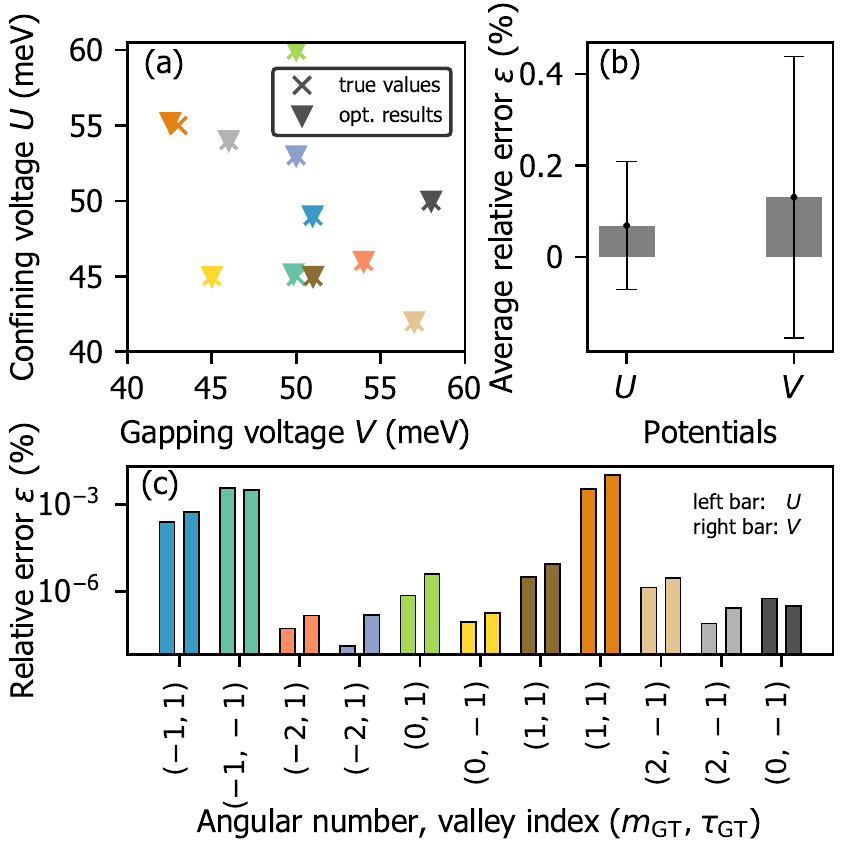}
    \caption{Testing of the HRS algorithm on the simulated data: We choose 11 different ground-truth states fully characterized by parameters ($m_{{\rm GT}},\tau_{{\rm GT}},U_{{\rm GT}},V_{{\rm GT}}$) and use HRS to search for optimal values $U_{{\rm opt}}, V_{{\rm opt}}$ for all chosen states, while considering $m$ and $\tau$ to be known. (a) Ground-truth values (crosses) and the resulting values from the optimizer (triangles) for $U$ and $V$. (b) The relative error of the  $U$ and $V$ optimization averaged over 11 independent test runs. (c) The relative error of $U$ and $V$ for each of the test runs, which average to the value present in panel (b).}
    \label{fig:HRS_test}
\end{figure}

\begin{figure}
    \centering
    \includegraphics[scale = 1]{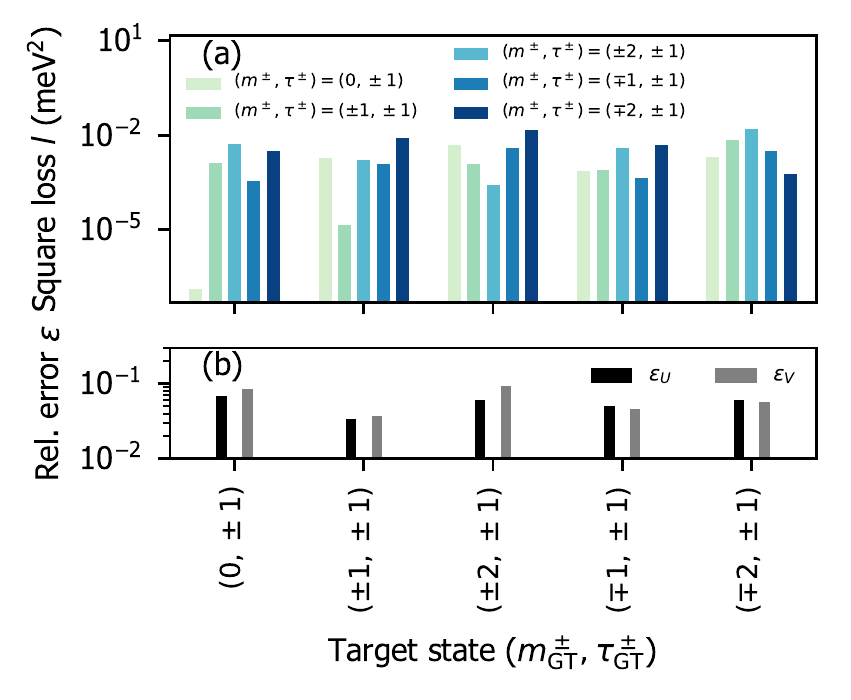}
    \caption{ (a) Mean
    squared error loss of five parallel runs for each of five target states characterized by the respective ($m_{\rm GT}, \tau_{\rm GT} $) pair shown on the x-axis. The five different bar colors correspond to the five best candidates tested.
    (b)~Re\-lative errors of confining and gapping potentials $U,V$ for the five target states.
    }
    \label{fig:control_benchmark}
\end{figure}

\begin{figure}[!t]
    \centering
    \includegraphics[scale = 1.0]{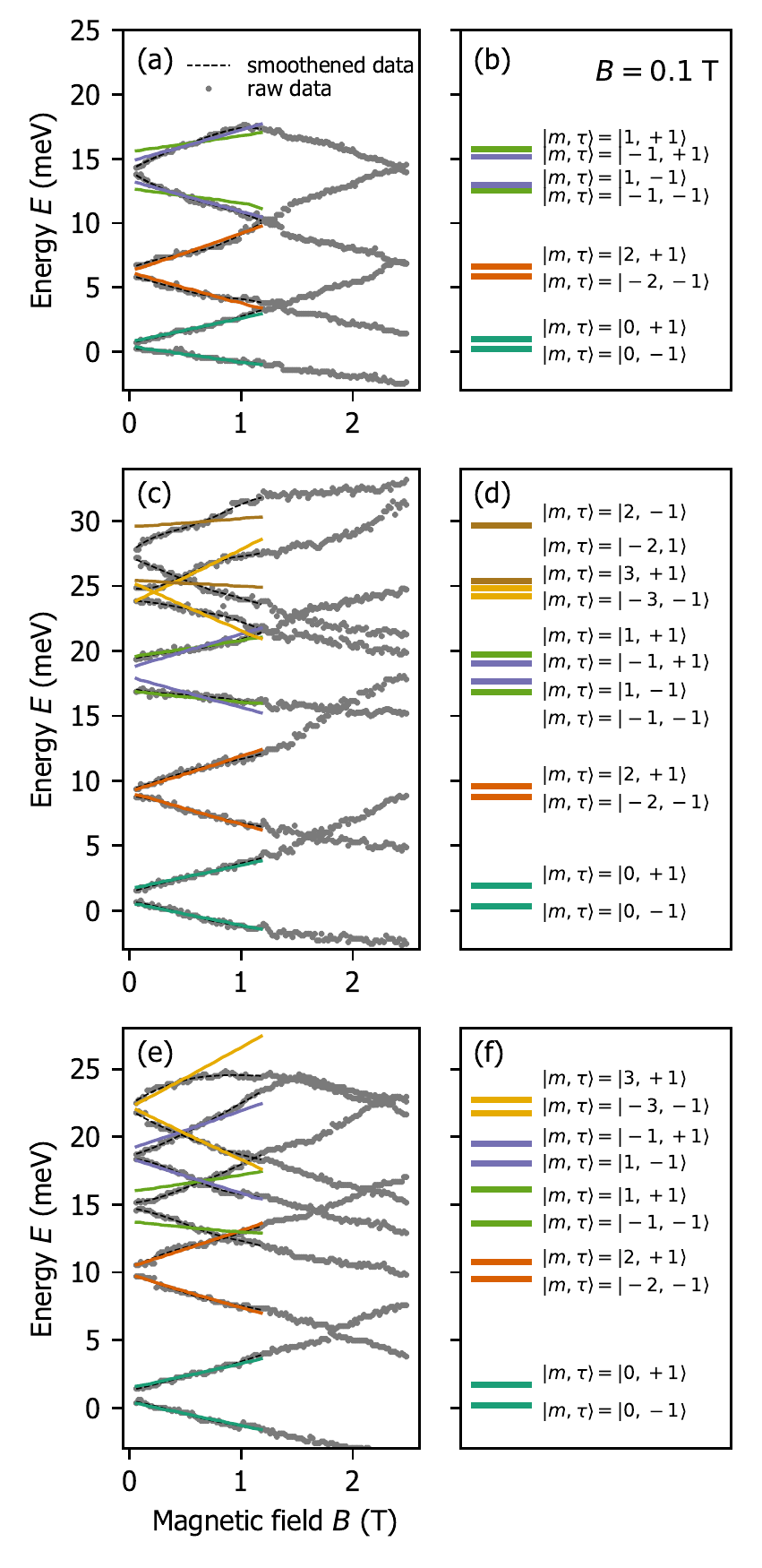}
    \caption{Performance of the HRS algorithm on the experimental data consisting of single-particle states for three different bilayer graphene quantum dot. In panels  (a), (c) and (e), grey dots represent the measured energy $E$ as a function of the modulated magnetic field $B$, dashed lines correspond to smoothened data and colored solid lines display best fits obtained up to $B=1.2$~T computed  by the HRS (and CRS-LM for the lowest-lying energy pair) algorithm. Panels (b), (d) and (f) illustrate the extracted single-particle energy ladders at $B=0.1$~T fully resolved by differing angular momenta $m$ and valley numbers $\tau$.}
    \label{fig:experiments_all}
\end{figure}

We now test the HRS algorithm on both numerically generated and experimentally measured data. 
The summary of results of the application of the method on the numerically generated data is shown in Fig.~\ref{fig:HRS_test}. To simulate the measurements based on the low-energy theory introduced in Sec.~\ref{sec:bi}, we compute the determinant maps as shown in Fig.~\ref{fig:TSNE}(a), (b) with respect to the Hamiltonian in Eq.~\eqref{eq:H0b_1} for 11 sets of parameters $\{U,V,m,\tau\}$ (see Fig.~\ref{fig:HRS_test}(a)) and extract the energies, $E$, for which the determinant is zero as a function of the magnetic field, $B$ [as shown by the dashed lines in Fig.~\ref{fig:TSNE}(a), (b)]. After applying HRS on this dataset, we calculate  the average relative error $\epsilon = \frac{1}{n}\sum_{j=1}^n \epsilon_j$ of the $U,V$ optimization shown in Fig.~\ref{fig:HRS_test}(b) as a mean of the relative errors $\epsilon_j = \vert Q_{\rm opt} - Q_{\rm GT}\vert /Q_{\rm GT}$ with $Q\in \{U,V\}$ of $n=11$ test optimization runs. We obtain the final value of $\epsilon = 2 \cdot 10 ^{-3}$ (or 0.2\%) for $U$ and $V$ estimation, while the maximum obtained error is 1\%. The error for each of the $n$ individual runs for specific fixed pairs of discrete quantum numbers $(m_{\rm GT},\tau_{\rm GT})$ is shown in Fig.~\ref{fig:HRS_test}(c).

As a comparable (but unguided by physics knowledge) random search benchmark, we choose Controlled Random Search with Local Mutation (CRS-LM). The CRS-LM approach yields an average error of $U$ and $V$ search reaching 2\% while the maximum error obtained is 8\%. The graphical summary of the CRS-LM results can be found in Fig.~\ref{fig:CRS-LM_test} in Appendix~\ref{app:further_res}. In addition, we observe that our Hamiltonian-guided ansatz HRS provides an order of magnitude precision improvement.

We need to address an important distinction between simulated and experimental data when applying HRS to the experimental data.
While the theoretical model introduced in Sec.~\ref{sec:bi} provides single-particle states on the absolute energy scale, for the experimental data the energy axis is determined relative to the lowest-lying state measured in the particular experimental realization. Thus, the absolute energy scale is not defined a priori. Therefore, when deploying the algorithm on experimental data, we first need to determine where the states lie on the absolute energy scale of the model. Then, we can launch the Hamiltonian search procedure described above.

We determine this scale by fitting the two lowest-lying ($\tau=\pm 1$) states for each quantum dot. For this task, we need to use a different approach as we are prevented from fitting the energy lines $E(B)$ directly without knowing the absolute energy scale. 
However, without the absolute energy scale, the previously explained benefit of incorporating physical knowledge into the optimization via the Hellman-Feynman theorem is no longer available to us. Hence, we apply the standard CRS-LM global search (which we previously used as a benchmark for HRS) on the gradient of the measured data and optimize parameters $U_{\rm GT},V_{\rm GT}$ of the theoretical model. Once the energy scale of the measured data is determined by fitting the parameters of the lowest-lying state pair with CRS-LM, we can fit all higher-lying states (of both simulated and experimental data) using our hybrid HRS approach the same way as shown above for simulated data.

Using the available estimate of experimentally relevant values, we narrow the $U-V$ domain for the lowest-lying pair to $50$-$70$ meV~\cite{eich_2018}. To obtain a sufficient resolution, we divide this domain into 25 subdomains and perform a global search via CRS-LM on each of them for all relevant combinations of $m$ and $\tau$. The use of CRS-LM leads to lower fitting precision for the lowest-lying states as opposed to the rest of the spectrum. 

As an illustration of the fitting of the lowest-lying state, we show the accuracy of the procedure for various discrete quantum numbers in Fig.~\ref{fig:control_benchmark} using simulated target states resulting from the theoretical model given in Eq.~\eqref{eq:H0b_1}. Figure~\ref{fig:control_benchmark}(a) shows the value of the loss function, Eq.~\eqref{eq:square_loss}, for a combination of the five best candidates of the discrete quantum number pairs $m$ and $\tau$ as a function of the respective ground-truth values used to compute the energies. We see that the smallest loss value after the optimization indeed corresponds to the ground-truth parameter pair in each case. It can be observed that for some ground truth states in Fig.~\ref{fig:control_benchmark} (namely target state ($m_{\rm GT}, \tau_{\rm GT}) = (\mp 1, \pm 1)$) the best candidate is not substantially better than the other candidates. To investigate the robustness of these minima, we carefully analyze the features in their immediate neighborhoods. We average over $M = 1000$ points of the small neighborhood of each identified minima. We identify no overlap in the standard deviation intervals by comparing mean values and their standard deviations. The differences in the loss function values while small are robust and well-separated.

Figure~\ref{fig:control_benchmark}(b) shows the relative estimation error of $U, V$ averaged over the discrete quantum numbers of the lower and upper state that appears as a consequence of the valley splitting. Specifically, $\epsilon_Q = \frac{1}{2}\sum_{s\in\{+,-\}} |Q_{\rm GT}^s - Q_{\rm opt}^s|/Q_{\rm GT}^s$. For the numerically generated data, we are able to estimate the continuous variables with a precision of $3-7~\%$ for the two lowest-lying states while reliably determining the discrete variables.

Above, we discussed the HRS method and its performance on the numerically generated data. Now, we move forward to apply the method on experimentally measured data. In our experimental realization, we have measured three separate bilayer graphene quantum dots (QD1, QD2, and QD3, respectively).
The performance of our algorithm applied on these experimental measurements from the individual quantum dot systems is shown in Fig.~\ref{fig:experiments_all}.

For the lowest-lying states, we consistently find $m_{\rm opt,0}^+,\tau_{\rm opt,0}^+ = (0,1)$ for the increasing and $(m_{\rm opt,0}^-, \tau_{\rm opt,0}^-) = (0,-1)$ for decreasing valley-splitted branch of the state across all three samples. Here, the index $0$ denotes the lowest-lying couple of the measured energy states that we use to fix the energy scale. For all the remaining states, we performed HRS considering each feasible combination of $m,\tau$ and domain $Q_i^s\in \{Q_{\rm opt, 0}^s - 5, Q_{\rm opt, 0}^s + 45\}$  meV where $i$ denotes the state pair index, $i\in\{1,...,N\}$. These domain boundaries are physically motivated by the measured energies range we observed. We use the mean squared error function, Eq.~\eqref{eq:big_loss}, to determine the optimal assignment of the discrete quantum numbers for the energy ladder.

\begin{figure}
    \centering
    \includegraphics{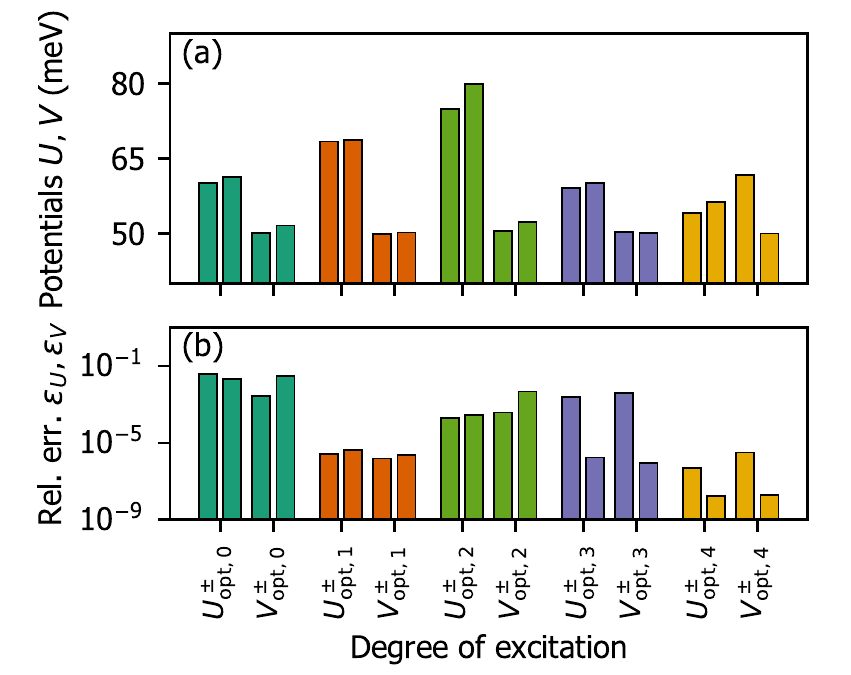}
    \caption{Values (a) and statistical errors (b) of confining $U$ and gapping  $V$ potentials for QD3 based on the HRS (and  CRS-LM  for  the  lowest-lying  energy  pair $i=0$) 
    optimization routine. Notation $Q_{\rm opt,i}^\pm$ comprises potential $Q\in\{U,V\}$ of increasing ($+$) and decreasing ($-$) spectral line from $i$-th spectral couple consisting of degenerated states at $B=0$~T.}    \label{fig:UV_values_errors_CG9}
\end{figure}

Let us specify how we compute the statistical accuracy of the estimation of the continuous parameters. In the case of two lowest-lying states and  CRS-LM search on 25 subdomains of the $U-V$ plane, we performed 10 parallel runs to find the optimal discrete parameters. 
For all the remaining states and HRS optimization, 5 parallel runs were executed for the best quantum number candidates (due to the increased number of possible parameter combinations and the results consistency,  we restricted ourselves to 5 parallel runs per combination). We then calculate the mean and standard deviation of these estimates via
\begin{align*}
    Q^s_{\rm opt,i} &= \frac{1}{p}\sum_{j=1}^p Q^s_{\rm opt_j,i}\\
    \epsilon_{Q^s_{\rm opt,i}} &= \sqrt{\frac{1}{p}\sum_{j=1}^p \left(Q^s_{\rm opt_j,i} - Q^s_{\rm opt,i}\right)^2}. 
\end{align*}
Here, $Q^s_{\rm opt_j,i}$ is the optimized value of the potential $Q^s\in\{U^s,V^s\}$ for the $i$-th spectral lines pair, the index $j$ denotes the respective parallel run, and the total number of the runs is denoted by $p$. 

In Fig.~\ref{fig:UV_values_errors_CG9}, we visualize the results for the estimation of $U$ and $V$ for QD3. Equivalent studies for the two remaining dots QD1 and QD2 can be found in Figs.~\ref{fig:UV_values_errors_CG2} and \ref{fig:UV_values_errors_CG3} in Appendix~\ref{app:further_res}.
We conclude that the estimation error for the confining and gapping potentials on experimental data has a statistical error of approximately $5\%$. 
In addition, we summarize the parameters of each fitted single-particle state together with the associated statistical errors of the continuous parameters for all three quantum dots in Tab.~\ref{tab:U_V_values_and_errors} in Appendix~\ref{app:further_res}.

\begin{figure}
    \centering
    \includegraphics{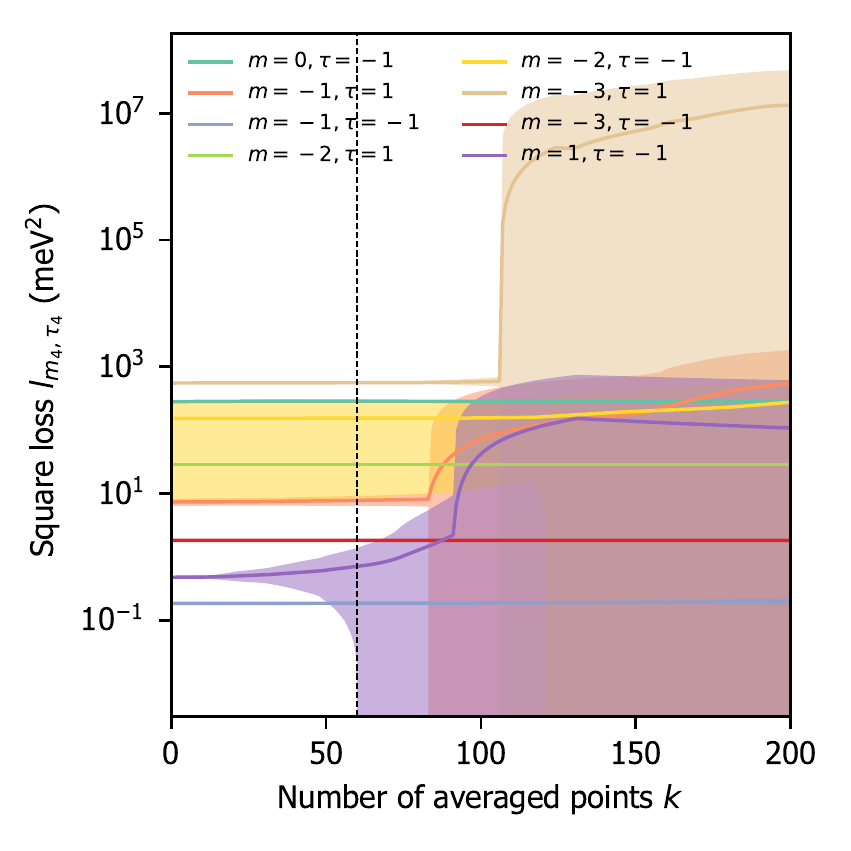}
        \caption{Robustness of the optimization procedure of the discrete numbers $m,\tau$ for higher-lying energy states visualized for the fourth (experimentally measured) spectral line pair.
        On the $x$-axis we show the number of smallest-loss $(U,V)$ configurations, $k$, which we average over when constructing the average square loss shown on the $y$-axis (mean value in solid, standard deviation as shaded regions). Different colours refer to the different  discrete quantum numbers assumed in the optimization process for potentials $U$ and $V$.
        }
    \label{fig:k_averages}
\end{figure}

Finally, we analyze the robustness of the discrete variable search by averaging over number of points, $k$, for which the loss function $l_{m,\tau}$ attains the smallest value. This is motivated by the fact that the measured data are subject to small fluctuations, and we want to avoid finding an optimum having such a non-systematic origin. Therefore, we need also to investigate several next-to-optimal configurations and compare them between respective ($m,\tau$) choices. Taking $k\rightarrow 0$ (approaching the optimal configuration) could not be sufficient to reliably eliminate such fluctuation effects while at too large $k$ we deviate significantly from the optimum found (the second can be seen as an abrupt increase of loss values for larger $k$ in Fig.~\ref{fig:k_averages}). Therefore, we experimentally choose $k=60$ to avoid both limiting cases described. 
We show the loss $l_{m,\tau}$ of all $(m,\tau)$ candidates for state pair $i=4$ and QD2 averaged over the $k$ best configurations as a function of $k$ in Fig.~\ref{fig:k_averages} (solid lines). 
The 2D-optimization in the $U-V$ plane is performed for each relevant combination of discrete parameters $m_4$ and $\tau_4$ and optimal continuous parameters $U^{s}_{{\rm opt},i=4},V^{s}_{{\rm opt},i=4}$ are found by minimizing Eq.~\eqref{eq:square_loss}. The square loss $l_{m_4,\tau_4}=\frac{1}{2}\sum_{s\in\{\pm\}} l_{m_4^s,\tau_4^s} \left(U^{s}_{{\rm opt},4},V^{s}_{{\rm opt},4}\right)$ refers to the loss attained after the optimization, i.e., the individual contributions to the total cost function of the inner sum in Eq.~\eqref{eq:big_loss}. The standard deviations of obtained $l_{m_4,\tau_4}$  are represented by light-shaded regions. We see robust differences between the best discrete variable candidate pairs as the loss averages stay well detached for a sufficient amount of averaged points.

\section{Discussion and future work}
\label{sec:disc}

Let us discuss the limitations and perspectives of our work. In the present analysis, we have used a minimal model, which is the simplest model at our disposal to capture the low-energy physics of bilayer graphene quantum dots. The following set of assumptions characterizes this model: i) The model relies on a physically motivated linearization of the dispersion relation in currently experimentally most relevant BGQD structures with AB stacking. ii) A tight-binding approximation. iii) Restrictions on the symmetry of the quantum dot. iv) Continuity of the wave function on the boundary of the quantum dot. v) No additional spin degree of freedom. vi) A description of electron bound state occupying the BGQD based on the solution of the one particle Schrödinger-like equation.

While the model does not explicitly include every aspect of bilayer graphene quantum dots, the choice of model has the consequence that all free parameters are readily interpretable, and a semi-analytical solution of the model exists. Therefore, we can generate large data sets that allow us to rigorously benchmark our method against other techniques at low computational cost (see Appendices~\ref{app:unsupervised} and~\ref{app:nn_discrete}). Despite its simplicity, we have found that our model can well describe the experimental data of three quantum dots by fitting its parameters. In particular, the energy gaps between and the slopes of low-energy bound states agree almost perfectly with the experimental data.

As expected from the assumptions, higher-lying excited states are not perfectly described by our model, as can be seen in Fig.~\ref{fig:experiments_all}. For example, the mismatch between our fit and the highest-lying energy state in Fig.~\ref{fig:experiments_all} (e) is caused by the fact that the energy state is approaching the gap boundary. One of the mentioned approximations made is the assumption about the dot symmetry and a straight forward extension of the model would be to include perturbations of the cylindrical shape of the dot. However, as we do not posses the information about the exact shape of the potential, one can include this as an additional free parameter to be fitted.

For future considerations, one can analyze the role of the non-dimer atom sites tunneling, causing trigonal warping or electron-electron exchange interactions \cite{knothe_2020, garreis_2021}, both neglected in our theory assumptions. In fact, it would be very interesting to combine the most recent available experimental data \cite{garreis_2021, moller_2021} together with the most advanced theory description \cite{knothe_2020, knothe_2022}, through the algorithmic methodology introduced here.

As for the methodology, an exciting direction for future research is the additional customization of the global optimizer used in our algorithm. Specifically, one could develop mutations within the CRS method tailored to the structure and physics of the problem at hand to decrease the number of test points the algorithm needs to consider and, therefore, further boost the efficiency of the optimizer.

\section{Conclusions}
\label{sec:conclusion}

We have introduced the hybrid optimization algorithm HRS to fit the Hamiltonian and wave function parameters from bilayer graphene quantum dot transport measurements.
We have tested our method on computer simulations and experimental measurements of bilayer graphene quantum dots and obtained consistent results. We are able to reconstruct the wave function and Hamiltonian parameters with the statistical error of approximately $5~\%$ in the case of experimental data and relative estimation error of $0.2\%$ ($3-7\%$) for excited (ground) state of the simulated data.

Our method opens an avenue towards precise wave function reconstruction from noisy experimental data, specifically in situations when the optimization landscape manifests a large number of local minima that are hard to distinguish. Such large number of local minima is the situation one frequently faces in the case of transport measurements used for the characterization of quantum devices. HRS incorporates the initial knowledge of the physics of the problem to radically confine the optimization search domain such that powerful gradient-free but computationally demanding global methods become feasible.

\section*{Acknowledgements}

We acknowledge fruitful discussions with Klaus Ensslin and Sebastian D. Huber. We are grateful for financial support from the Swiss National Science Foundation, the NCCR QSIT, as well as for financial support provided by the Slovak Research and Development Agency under Contract No. APVV-19-0371, by the agency VEGA under Contract No. 1/0640/20 and by the European Union's Horizon 2020 research and innovation programme under the Marie Sk{\l}odowska-Curie grant agreement No. 945478. This work has also received funding from the European Research Council under grant agreement no. 771503.
\begin{appendix}

\section{Eigenenergies and eigenstates for axially symetric bilayer graphene quantum dots}\label{app:model_analytical}

In this Appendix, we provide details regarding the zero-determinant solution introduced in Ref.~\cite{recher_2009}. The Brillouin zone of the bilayer graphene quantum dot system has four sites, which we label $A1,A2,B1,B2$ (with $A$ and $B$ referring to different graphene layers and we assume Bernal stacking with $B1,A2$ being the closest interlayer sites). We define the full 4-spinor
\begin{equation}
\Psi = 
  \begin{pmatrix}
    \Psi_{B1}\\
    \Psi_{A1}\\
    \Psi_{A2}\\
    \Psi_{B2}
  \end{pmatrix}
\end{equation}
of the bilayer graphene quantum dot system. Due to axial symmetry of the system at hand, it is natural to work in polar coordinates ($r,\varphi$) in which one can easily factorize the 4-spinor into radial and angular part.  In what follows we define $\Psi_1$ as

\begin{equation}
\label{eq:transform1_app}
\Psi(r,\varphi) = 
\frac{e^{ i m \varphi}}{\sqrt{r}}
\begin{pmatrix}
1 & 0 & 0 & 0 \\
0& e^{- i \varphi} & 0 & 0 \\
0& 0 & 1 & 0 \\
0 & 0 & 0 & e^{ i \varphi}
\end{pmatrix}
\Psi_1(r).
\end{equation} 
Therefore, $\Psi_1$ includes most of radial dependence of original spinor $\Psi$. Additionally, we denote $\Psi_2$ to be a vector related to the radial spinor part $\Psi_1$ via  
\begin{equation}
\label{eq:transform2}
\Psi_1(r) =
\begin{pmatrix}
\phi_{m}^s & 0 & 0 & 0 \\
0& \phi_{m-1}^s & 0 & 0 \\
0& 0 & \phi_{m}^s & 0 \\
0 & 0 & 0 & \phi_{m+1}^s
\end{pmatrix} \Psi_2.
\end{equation}
Radial dependence in above formula is present in $\phi$ functions and thus $\Psi_2$ depends only on the model parameters. $\phi^s_m$ functions are proportional to confluent hypergeometric (Kummer's) functions and apart from the dependence on angular momentum $m$, they bear the information about the orientation of a perpendicular magnetic field with respect to the graphene layers $s \in \{\pm1\}$.  

Thanks to a complete set of commuting observables and thus the existence of a common eigenbasis, we may solve the first order equation using Kummer's functions, to cast the problem into the shape of following homogeneous equation
\begin{equation}
A \Psi_2 = 0,
\label{eq:homogeneous-APsi2=0}
\end{equation}
where the 4 x 4  matrix $A$ reads
\begin{equation}
\small
A=\begin{pmatrix}
\frac{\tau V}{2}-\epsilon(r) & -i a_{1}^{s}/\sqrt{2}l_{B}& t_\perp & 0 \\
- i a_{2}^{s} /\sqrt{2}l_{B}& \frac{\tau V}{2}-\epsilon(r) & 0 & 0 \\
t_\perp & 0 & -\frac{\tau V}{2} -\epsilon(r) & - i a_{3}^{s}/\sqrt{2}l_{B} \\
0 & 0 & -i a_{4}^{s}/\sqrt{2}l_{B} & -\frac{\tau V}{2} -\epsilon(r)
\end{pmatrix}.
\label{eq:4x4forPsi2}
\end{equation}
Here, $\epsilon(r) = E-U(r)$.
Evidently, to have a nontrivial $\Psi_2$, the determinant of $A$ in Eq.~\eqref{eq:homogeneous-APsi2=0} must be zero, such that the
4-spinor can be  straightforwardly computed. However, we have to fix two remaining free parameters, namely  $\kappa$ defined as $\mathcal{H}_0 \Psi_1 = -i \kappa \Psi_1/\sqrt{2}l_B$ entering the model equations via $a^s_i$ functions and $E$. The condition on the singularity of $A$ yields the following relation between $\kappa$ and $E$:
\begin{eqnarray}
\label{eq:kappas}
\frac{\kappa_{<,>}^2}{2l_{B}^2}&=&\frac{s}{l_{B}^2}-\epsilon_{<,>}^{2}-\frac{V^2}{4} \nonumber\\
&\pm&\sqrt{t_\perp^{2}\left(\epsilon_{<,>}^{2}-\frac{V^2}{4}\right)+\left(\epsilon_{<,>}\tau V-\frac{s}{l_{B}^{2}}\right)^2}\,,
\end{eqnarray}
where we introduced the notation $<,>$ to distinguish quantities inside and outside of the dot. Let us suppose 
\begin{equation}
  U(r) = \begin{cases}
      0, &r \leq R \\
      U_0, &r>R 
      \end{cases}
\end{equation}
and, therefore, $\epsilon_< = E, \epsilon_> = E-U_0$ which fixes the parameter $E$. The continuity of the 4-spinor at the boundary of the dot allows us to fix the remaining degree of freedom, $\kappa$.
In both cases (inside and outside), we have two values of $\kappa$ and thus also two solutions $\Psi_2^+$ and $\Psi_2^-$. This implies that the full solution inside $\Psi_{1,<}$ and outside $\Psi_{1,>}$ the dot reads
\begin{align}
\label{eq:psi_1_solIn}
\Psi_{1,<} &= A \Psi_{1,<}^+ + B \Psi_{1,<}^-\,, \\
\label{eq:psi_1_solOut}
\Psi_{1,>} &= -C \Psi_{1,>}^+ -D \Psi_{1,>}^-\,.
\end{align}
We find the energy of the respective state $\Psi_1$ by matching these two states from Eq.~\eqref{eq:psi_1_solIn} and \eqref{eq:psi_1_solOut} at $r = R$. This finally corresponds to solving for the zero determinant of
\begin{gather}
\label{eq:psi1_det}
A(E| r=R) = \left(
  \centering
  \begin{tabular}{c|c|c|c}
  \multirow{4}{*}{$\Psi_{1,<}^+(E)$} & \multirow{4}{*}{$\Psi_{1,<}^-(E)$} & \multirow{4}{*}{$\Psi_{1,>}^+(E)$} & \multirow{4}{*}{$\Psi_{1,>}^-(E)$}\\
  & & &\\
  & & &\\
  & & &\\
  \end{tabular}
\right)
\end{gather}

\section{Unsupervised learning}\label{app:unsupervised}

In Sec.~\ref{sec:bi}, we have demonstrated the extraction of the discrete quantum numbers (angular momentum $m$ and valley number $\tau$) based on numerically generated 2D determinant maps concerning bilayer graphene quantum dots described by Hamiltonian~\eqref{eq:H0b_1}. To this end, we have used the t-distributed Stochastic Neighbor Embedding (t-SNE) \cite{maaten_2008_tsne_main}, which is a non-parametric visualization and dimensional reduction technique. In this technique, a spatial distribution of data points in a high-dimensional space is modeled by Gaussian kernels, whereas the positions of the data points embedded in a low-dimensional space are expressed via Student-t kernels. The latter distribution has a heavy tail, which helps to account for far separated points in the high-dimensional space. The low-dimensional probability distribution is then found by minimizing the Kullbeck-Leibler divergence between the spatial distributions of the data points in the high- and low-dimensional space. For more details, we refer the reader to \cite{maaten_2008_tsne_main} and references therein.    

The results of t-SNE applied on the bilayer graphene quantum dot model data are shown in Fig.~\ref{fig:TSNE}(c) in Section~\ref{sec:bi}. Figure~\ref{fig:TSNE}(c) contains altogether 4410 (21 values for both $U,V$ from range 50-70 meV with step 1 meV for 10 combinations of $m$ and $\tau$) data points that are well separated in ten distinct classes defined by unique combinations of $m$ and $\tau$ values. Each data point represents a single 2D determinant map provided by the bilayer graphene quantum dot model for fixed parameters $m,\tau,U$ and $V$. We use a resolution of  of $360\times360=129600 \,{\rm px}$ for the maps. We determined this resolution by the trade-off between having a smooth and well-defined determinant profile and computational feasibility. 

Nevertheless, the application of t-SNE, as outlined above, is not appropriate as a convenient pipeline for an efficient prediction of the ground-truth discrete quantum numbers, as t-SNE is not re-usable for changing input data. To circumvent this computational issue, we introduce kernel Principal Component Analysis (kPCA) as a pre-processing step. In contrast to t-SNE, kPCA is a parametric method, and thus after training, kPCA can be directly re-applied on new input data. Therefore, we combine t-SNE with kPCA to engineer a much faster tool, as compared with t-SNE on its own, while showing an almost identical clustering as in Fig.~\ref{fig:TSNE}(c).

\begin{figure}
    \centering
    \includegraphics{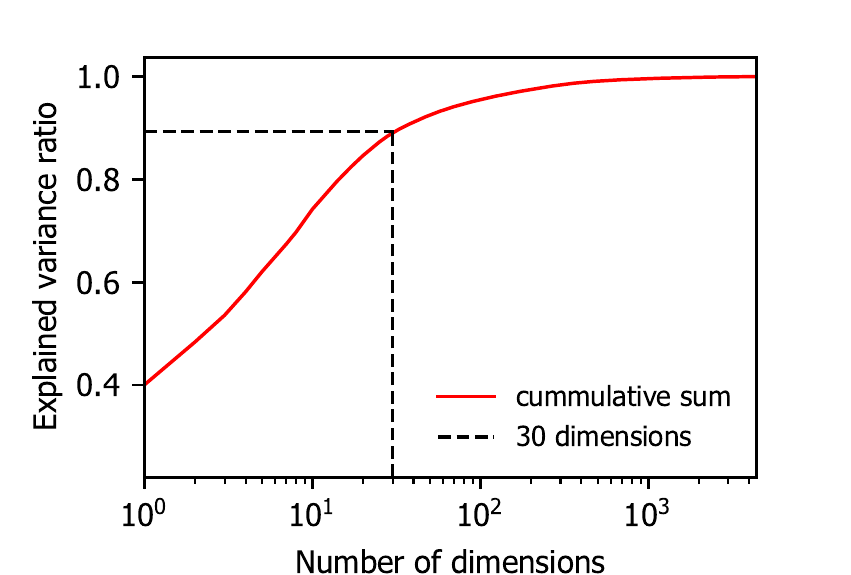}
    \caption{Cumulative sum of explained variance ratio for kernel PCA (red) applied on 4410 determinant maps of the bilayer graphene quantum dot model. The chosen cut-off of 30 principal components corresponding to $\approx90\%$ of explained variance is indicated by the black dashed line.}
    \label{fig:cum_sum}
\end{figure}

The dramatic reduction of computational effort may be attributed to the small number of required principal components in kPCA in our setting. Investigating the explained variance ratio of kPCA applied on our generated data set, we observe that 30 principal components are sufficient to capture 90\% of the data variance as shown in Fig.~\ref{fig:cum_sum}. Therefore, once the weights of kPCA are trained to map the data of the 129600-dimensional space into the reduced 30-dimensional space (while keeping most of the variance), t-SNE may be efficiently used also on new data after the application of the kPCA module. 

Based on the combined usage of kPCA and t-SNE for the prediction of $m,\tau$, it is conceptually straightforward to estimate the remaining continuous parameters $U$ and $V$ based on pre-trained CNNs as described in Appendix~\ref{app:nn_det}. Thus, this represents an alternative, sequential routine for the accurate prediction of all parameters of the bilayer graphene quantum dot Hamiltonian~\eqref{eq:H0b_1}.   

\section{Direct usage of neural networks for the prediction of the continuous Hamiltonian parameters}
\label{app:nn_det}

In the main text, we demonstrated that we can retrieve the discrete quantum numbers, $m_{\rm GT}$ and $\tau_{\rm GT}$, from the numerically generated 2D determinant maps by the usage of the t-SNE algorithm, see also Appendix \ref{app:unsupervised}. However, we also showed that the experimental data (or the plain energy lines) do not contain a sufficient amount of information, such that $m_{\rm GT}$ and  $\tau_{\rm GT}$ can be inferred. Thus, we introduced another approach to extract  both discrete and continuous parameters in what followed. For completeness, we show in this section that the continuous parameters $U,V$ may also be inferred from 2D determinant maps based on convolutional neural networks (CNNs).

\begin{figure}[!htb]
    \centering
    \includegraphics[scale = .29]{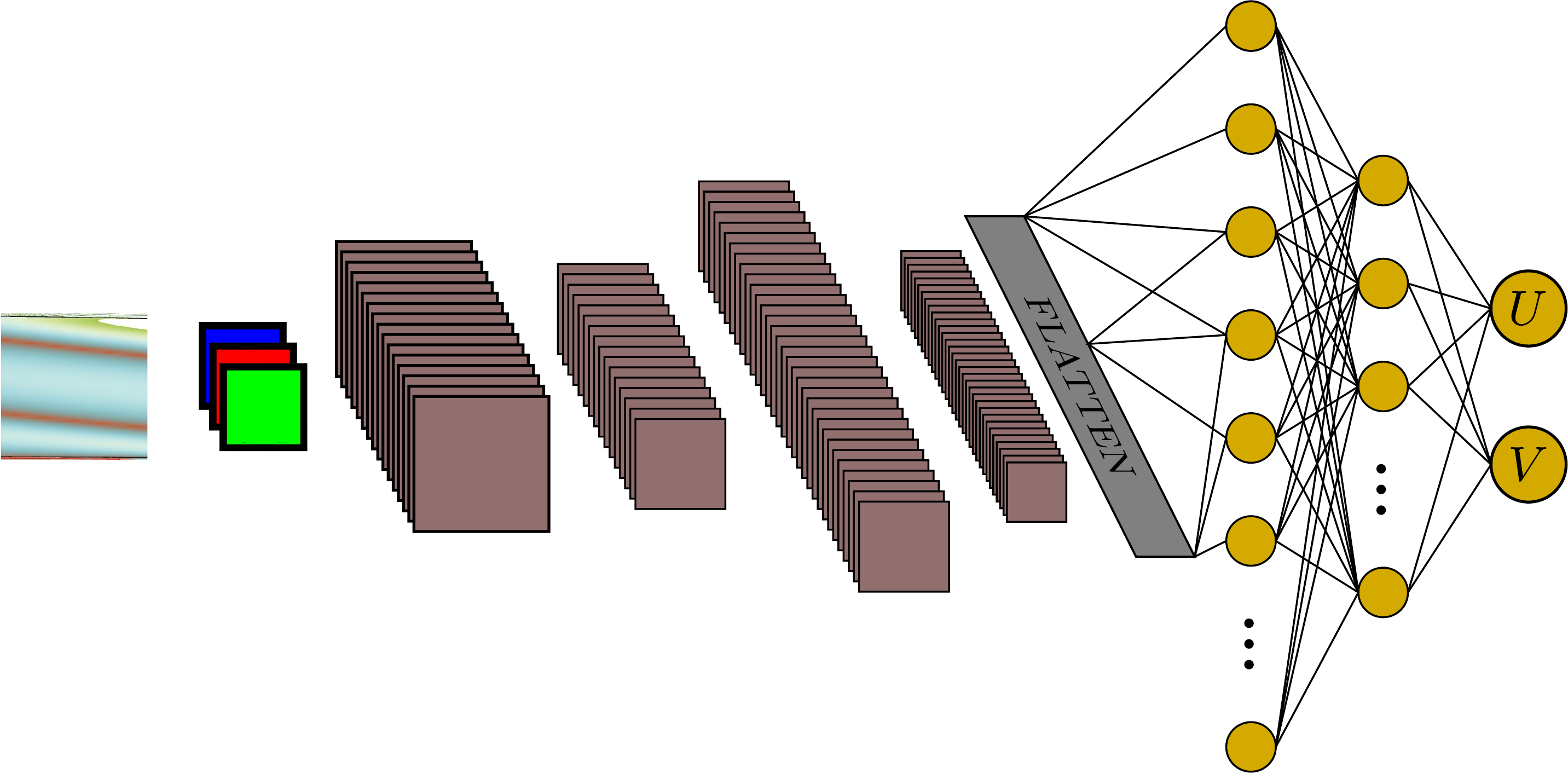}
    \caption{Sketch of the CNN architecture used for the prediction of the continuous parameters based on 2D determinant maps. Employed hyperparameters are listed in Tab.~\ref{tab:CNN_params}.}
    \label{fig:CNN_scheme}
\end{figure}

In this regard, we trained an individual CNN, but with fixed architecture, for each considered combination of quantum numbers $m,\tau$ in a supervised manner with determinant maps labeled by their ground truth values of $U_{\rm GT},V_{\rm GT}$. The CNN architecture is sketched in Fig.~\ref{fig:CNN_scheme}, which takes a determinant map as an input and outputs estimates $U_{\rm opt}, V_{\rm opt}$ of the continuous parameters $U$ and $V$. The associated hyperparameters of the CNN are summarized in Tab.~\ref{tab:CNN_params}.

\begin{figure}[!htb]
    \centering
    \includegraphics[scale = 1]{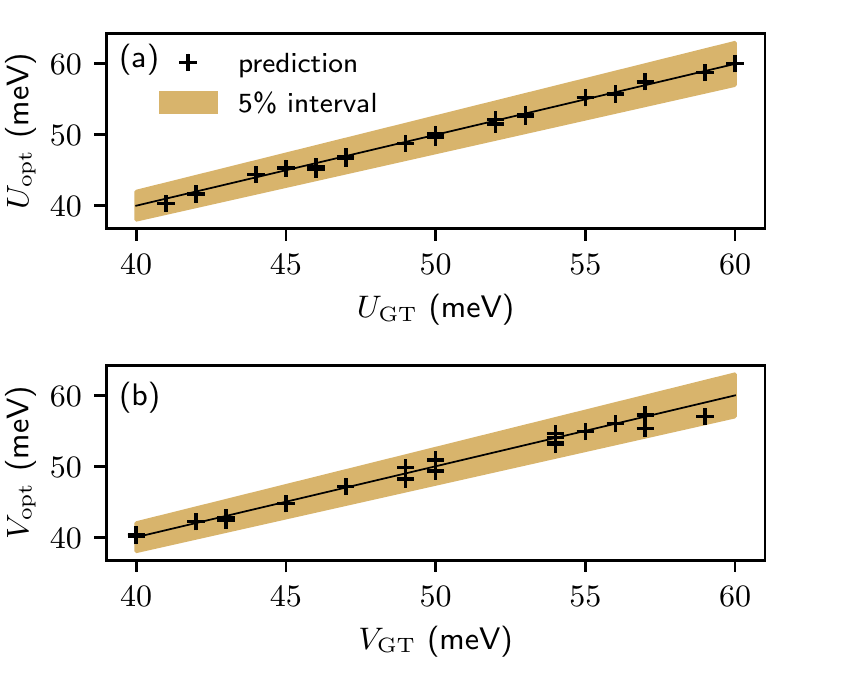}    
    \caption{Application of the trained CNN to predict confining (a) and gapping potential (b) in the case of fixed discrete quantum numbers $m=0$ and $\tau = 1$. The Solid black line represents the identity function depicting the ideal predictions. The brown shaded region shows the 5~\% relative error interval with respect to the ground truth value.}
    \label{fig:CNN_UV_predictions}
\end{figure}

Based on the t-SNE clustering as described in Sec.~\ref{sec:bi} and the resulting values of the discrete parameters, we select the respective CNN to predict the values of $U$ and $V$.   
We plot the predictions of the CNN based on fixed values of $m=0$ and $\tau = 1$ in Fig.~\ref{fig:CNN_UV_predictions}. We conclude that both continuous parameters can be inferred with about 1\% precision using this approach based on CNNs.

\begin{table}[!htb]
    \centering
    \begin{tabular}{|l |l|}
    \hline
    Input channels  & 3 (RGB, each $360\times 360$ px) \\
    \hline
    \multirow{5}{*}{$1^{\text{st}}$ convolution} & feature maps: 16 \\
       & kernel: $5 \times 5$ \\
       & stride: 1 \\
       & padding: $(2,2)$ \\
       & activation: ReLU\\
    \hline 
    \multirow{3}{*}{$1^{\text{st}}$ pooling} & kernel: $5 \times 5$ \\
       & stride: $(5,5)$ \\
       & type: max pooling\\
    \hline   
    \multirow{5}{*}{$2^{\text{nd}}$ convolution} & feature maps: 32 \\
       & kernel: $3 \times 3$ \\
       & stride: 1 \\
       & padding: $(1,1)$ \\
       & activation: ReLU\\
    \hline   
    \multirow{3}{*}{$2^{\text{nd}}$ pooling} & kernel: $3 \times 3$ \\
       & stride: $(3,3)$ \\
       & type: max pooling\\
    \hline   
    \multirow{3}{*}{$1^{\text{st}}$ dense} & type: linear \\
       & size: 250 \\
       & activation: ReLU\\
    \hline   
    \multirow{3}{*}{$2^{\text{nd}}$ dense} & type: linear \\
       & size: 50 \\
       & activation: ReLU\\
    \hline   
    \multirow{3}{*}{Output layer} & type: linear \\
       & size: 2 \\
       & activation: None\\
    \hline
    Dropout & before $1^{\text{st}}$ dense ($p=0.5$)\\
    \hline
    \multirow{2}{*}{Optimizer} & type: Adam\\
     & learning rate: 0.001\\
    \hline 
    Loss & Mean squared error   \\
    \hline
    Epochs & 150\\
    \hline
    Batch size & 100\\
    \hline
    \end{tabular}

    \caption{Hyperparameters of the CNN used for the inference of the continuous Hamiltonian parameters (see Fig.~\ref{fig:CNN_scheme} for a sketch of the CNN).}
    \label{tab:CNN_params}
\end{table}

\section{Experimental data and energy measurements pre-processing}\label{app:exp_data}

\begin{figure}[!htb]
    \centering
    \includegraphics[scale = 1]{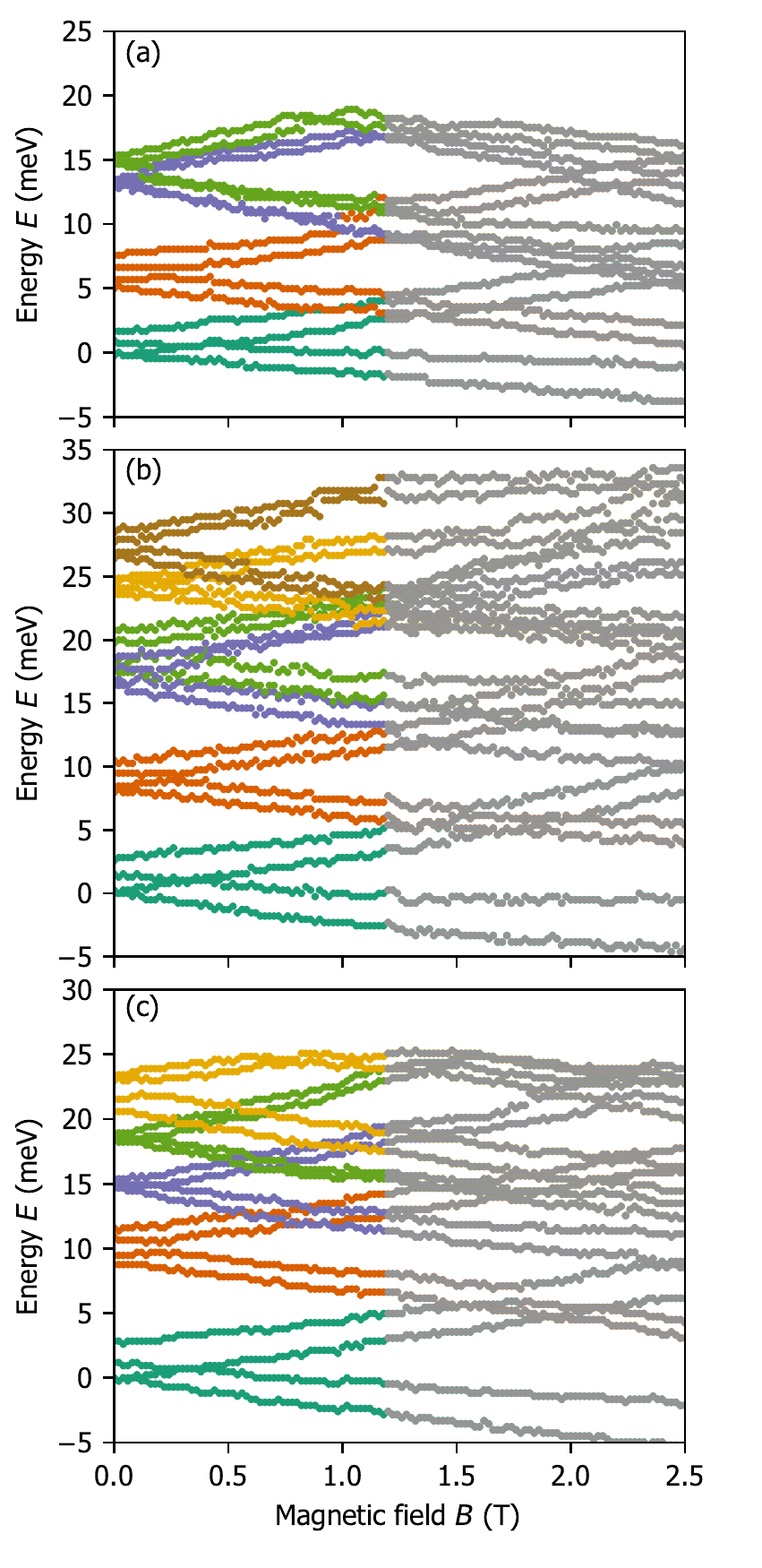}
    \caption{Experimental data as obtained from the transport measurements on three quantum dot systems QD1, QD2, and QD3. (a) Single-particle energy lines as a function of perpendicular magnetic field $B$ for QD1 system. Panels (b) and (c) show the same quantity for QD2 and QD3, respectively. The line coloring corresponds to the averaged energy lines displayed in Fig.~\ref{fig:experiments_all}. Gray-colored data above $B=1.2$~T are not used in our fitting.}
    \label{fig:exp_data_app}
\end{figure}

In this section, we present the experimental data used in our study. The experimental design of transport measurements was briefly discussed in Sec.~II of the main text. From the measured data, we extract the energy dependence on the magnetic field perpendicular to the graphene sheets for three different quantum dot systems QD1, QD2, and QD3 as shown in panels (a), (b), and (c) of Fig.~\ref{fig:exp_data_app}, respectively. 

As we discuss in the main text, our theoretical model based on Hamiltonian \eqref{eq:H0b_1} does not incorporate the spin degree of freedom that, except for the valley degeneracy, is responsible for additional two-fold degeneracy of the measured energy states \cite{kurzmann2019excited}. Altogether, we observe quadruplets of energy lines converging to a similar energy value as $B\rightarrow 0$. As different valleys can be clearly distinguished by looking at the overall line shape (increasing lines originate from one valley and decreasing from the other one), one can easily target the degenerate lines due to the spin degree of freedom. To account for the fact that we do not distinguish spins within our framework, we average over the spin degree of freedom. Thus, we effectively reduce the number of lines being fitted by a factor of 2. 

However, there is one caveat in the described procedure. We have to assign the measured spectral lines into the quadruplets degenerate by valley and spin freedom. This is an easy task once the energy scales within and between the lines of proposed quadruplets at zero magnetic field are different [e.g., see Fig.~\ref{fig:exp_data_app}(c)], in other words, if the quadruplets are significantly detached from each other. On the other hand, there are two cases in our data, where the distinction is not clear enough [purple and green lines in panels (a) and (b) of Fig.~\ref{fig:exp_data_app}]. In such a case, we end up with four increasing and four decreasing lines without a clear way of composing two quadruplets out of them. Therefore, in these two cases, we average over all four increasing and decreasing lines bearing in mind that we need to allocate twice as many discrete (and also continuous) Hamiltonian parameters to the averaged lines and thus provide two fits for such lines with two different fitting parameter sets. This explains why the purple and green fitted lines in panels (a) and (b) of Fig.~\ref{fig:experiments_all} belong to only a single experimental data line pair. At the very end of pre-processing, we perform data smoothing using Savitzky-Golay filter \citep{sav_gol_1964} to reduce the effects of experimental noise.

\section{Technical details of the HRS algorithm}

\subsection{Controlled Random Search with Local Mutation}
\label{app:crs}

Controlled Random Search (CRS) is an instance of a global optimization algorithm. Several variants of the algorithm have been proposed and benchmarked on different test problems with respect to other global optimizers~\cite{Arnoud2019BenchmarkingGO}. Inspired by the findings of Ref.~\cite{Arnoud2019BenchmarkingGO}, we adapt controlled random search with local mutation (CRS-LM)~\cite{kaelo_ali_2006_crs-lm} for our purposes. In what follows, we provide a (rough) step-by-step description of CRS-LM, as implemented in the NLopt package~\cite{johnson_nlopt}.

Let us assume an $n$-dimensional parameter space subjected to an optimization problem with loss function $l$. Then, the algorithm consists of the following steps:

\begin{enumerate}
    \item \textit{Initialization}: \\
    $N\gg n$ initial points are randomly chosen from the $n$-dimensional optimization domain (in our case $n=2$). Specifically, $N = 10(n+1)$ are chosen in the adapted implementation.
    \item \label{ranking} \textit{Points ranking}:\\
    Let $\mathcal{S}$ denote the set consisting of $N$ points as defined above. The points in $\mathcal{S}$ are ranked according to their loss value and let $x_b$ be the best point in the set (with smallest loss value $l$) and $x_w$ the worst point in the set (with largest loss value $l$).
    \item \label{new_point} \textit{Generating the trial point}:
    \begin{enumerate}
        \item \label{select} Select $n$ points  $x_2,x_4,...,x_{n+1}$ from $\mathcal{S}$ at random and take $x_1 = x_b$. The chosen points span a simplex in the $n$-dimensional space and the new trial point $\Tilde{x}$ is defined as a reflection of the last point $x_{n+1}$ through the centroid of $n$ remaining points
        \begin{equation}
            \Tilde{x} = 2 G - x_{n+1},
        \end{equation}
        with $G = 1/n\sum_{i=1}^n x_i$.
        \item If the trial point falls outside the optimization domain, return to \ref{select}.
        \item If the trial point $\Tilde{x}$ is worse than the worst point in $\mathcal{S}$ [$l(\Tilde{x})>l(x_w)$], go to \ref{mutation}, otherwise go to \ref{replace}.
    \end{enumerate}
    \item \label{mutation} \textit{Local mutation}:
    \begin{enumerate}
        \item The trial point chosen in step \ref{new_point} is worse than $x_w$ but instead of being discarded (as in standard CRS), here a local mutation is performed by coordinate-wise reflection of the trial point through the best point $x_b$. This reflection is defined as
        \begin{eqnarray}
            \Tilde{y}_i = (1+\omega_i)x_{bi} -\omega_i \Tilde{x}_i,
        \end{eqnarray}
        where $i$ denotes $i$-th coordinate and $\omega_i$ is chosen randomly from interval $[0,1]$.
        \item If $\Tilde{y}$ is not better than the worst point in $\mathcal{S}$ [$l(\Tilde{y})> l(x_w)$], no replacement is performed and the algorithm returns to \ref{new_point}.
         
    \end{enumerate}
        
    \item \label{replace} \textit{Update of $\mathcal{S}$}:\\ 
    Conditioned on whether coming from \ref{new_point} or \ref{mutation}, $\Tilde{x}$ or $\Tilde{y}$ takes place of the worst point $x_w$ in $\mathcal{S}$ (thus the actual $x_w$ is replaced by $\Tilde{x}$ or $\Tilde{y}$). The set is updated and algorithm returns to step~\ref{ranking}.
    
    \item Steps~\ref{ranking} - \ref{replace} are iterated until a stopping criterion is fulfilled. In our case, the algorithm was stopped once the relative tolerance in changes of the loss between two subsequent iterations was smaller than $10^{-4}$. This choice may in general be very problem-specific. For example, other common choices are the absolute tolerance in changes of the parameter vector or limiting the number of loss function calls.
    
\end{enumerate}

\subsection{The (hybrid) HRS algorithm}
\label{app:hrs}
In this section, we describe the HRS algorithm to infer the continuous parameters $U$ and $V$ of Hamiltonian \eqref{eq:H0b_1}. Let us suppose that the target state denoted by $U_{\rm GT},V_{\rm GT}$ is always chosen from the confined optimization domain \begin{equation}
   \mathcal{D} = (U_{\rm min},U_{\rm max}) \times (V_{\rm min},V_{\rm max})\,.
   \label{eq:optimization_domain}
\end{equation}
We introduce the HRS algorithm as a combination of a local and a global optimization algorithm to leverage the structure of the optimization landscape. 

\begin{enumerate}
    \item \label{gradient_descent} \textit{Gradient descent:}\\
       First, we aim at finding the narrow loss valley characterizing the bilayer graphene quantum dot optimization landscape, see Fig.~\ref{fig:landscape_transform}(a), by means of standard local optimization algorithms.
       We find that plain gradient descent is sufficient to identify this valley. We compute the required gradient by using the Hellmann-Feynman theorem, c.f. Sec.~\ref{sec:HRS}. 
       The gradient descent is terminated when the relative loss improvement between two consecutive epochs is under $2\%$.
       Ultimately, the (approximate) direction of the valley is defined by two points $P_1,P_2$ from two distinct gradient descent runs, which form a line specifying the direction of this confined loss region. In accordance with the relevant literature, we refer to the distinct gradient runs as ``walkers".
       
     \item \label{initialization_walkers} \textit{Convergence criteria \& initialization of the walkers:}\\   
     To have a meaningful approximation of the direction, we ensure that
       \begin{itemize}
           \item the two walkers do not end up too close to each other. Otherwise, a small error in the walkers' final position results in a large misalignment between the predicted and the actual valley direction. 
           \item if a walker terminates prematurely, i.e., without reaching the valley, a new walker is launched. 
           \item a walker remains within the optimization domain $\mathcal{D}$~\eqref{eq:optimization_domain}.
       \end{itemize}
       Therefore, we define a set $\mathcal{I}$ of candidate initial positions of the walkers:
        \begin{equation*}
        \mathcal{I} = \{c_1,c_2,c_3,c_4,s_1,s_2,...,s_{n_s}\},
        \end{equation*}
       where $c_1 \textnormal{-} c_4$ are the corners of the full rectangular optimization domain $\mathcal{D}$~\eqref{eq:optimization_domain} and $s_1 \textnormal{-} s_{n_s}$ are (pseudo) random inner points of this domain. 
       We use Sobol sequences (implemented in Ref.~\cite{sobol_seq}), which cover the landscape more efficiently than purely random distributed numbers. Then, $n_s$ denotes the length of the Sobol sequence which we set to $n_s = 15$.
       
       When the energy line  $E(B)$ touches the gap boundary or crosses it, i.e., when $E(B)$ is not entirely placed within the gap, a walker will terminate prematurely, because the energy dependence on the magnetic field is not well defined in such a case. Consequently, we restart the gradient descent on a new position of the set $\mathcal{I}$. 
        
    \item \label{descent_succesful} \textit{Determination of the confined (valley) domain}\\
        Once two walkers fulfill the criteria of step~\ref{initialization_walkers}, i.e., the two points $P_1,P_2$ have been determined and the direction of the valley may be estimated, as explained in step~\ref{gradient_descent}, then, the restricted optimization region (used later within the global optimization algorithm) is defined as a rectangle. The dimensions of this rectangle are given by the line segment between $P_1$ and $P_2$ and a margin of 4 meV [2 meV to each side of the valley line, see the illustration in Fig.~\ref{fig:landscape_transform}(b)].
        
        If there are more than $n_s + 4 = 19$ gradient descent attempts without success to determine the valley direction, the original domain $\mathcal{D}$~\eqref{eq:optimization_domain} is used for the global optimization algorithm.

    \item \textit{Transformation of the optimization domain: }\\
        The adapted global optimization routine \cite{johnson_nlopt} uses domains aligned with coordinate axes. Thus, we rotate the optimization domain, see Fig.~\ref{fig:landscape_transform} (b), in the case that the confined valley is found accordingly. 
    \item \textit{Global optimization: }\\
        Finally, we use CRS-LM to find the Hamiltonian parameters, $U$ and $V$, over the respective (confined) optimization domain.
\end{enumerate}

Our code for the HRS algorithm is open source~\cite{BGQD_code}.

\begin{figure}
    \centering
    \includegraphics{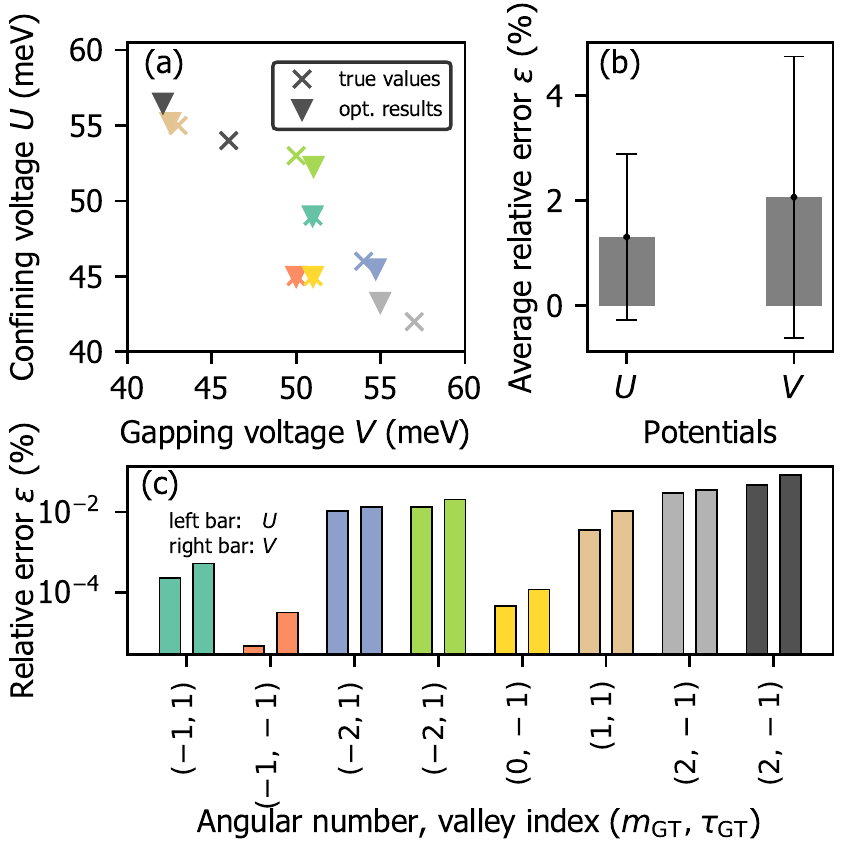}
    \caption{Testing of the CRS-LM algorithm on the numerically generated target states from Hamiltonian~\eqref{eq:H0b_1}. (a) Target states (crosses) and computed continuous parameters (triangles) in the $U-V$ plane. (b) Relative average error of the associated continuous parameters $U$ and $V$. Panel (c) depicts how the relative errors are distributed among the discrete quantum numbers of the respective target state.}
    \label{fig:CRS-LM_test}
\end{figure}

\section{Fitting parameters for additional quantum dot measurements}
\label{app:further_res}

\begin{figure}[!tb]
    \centering
    \includegraphics{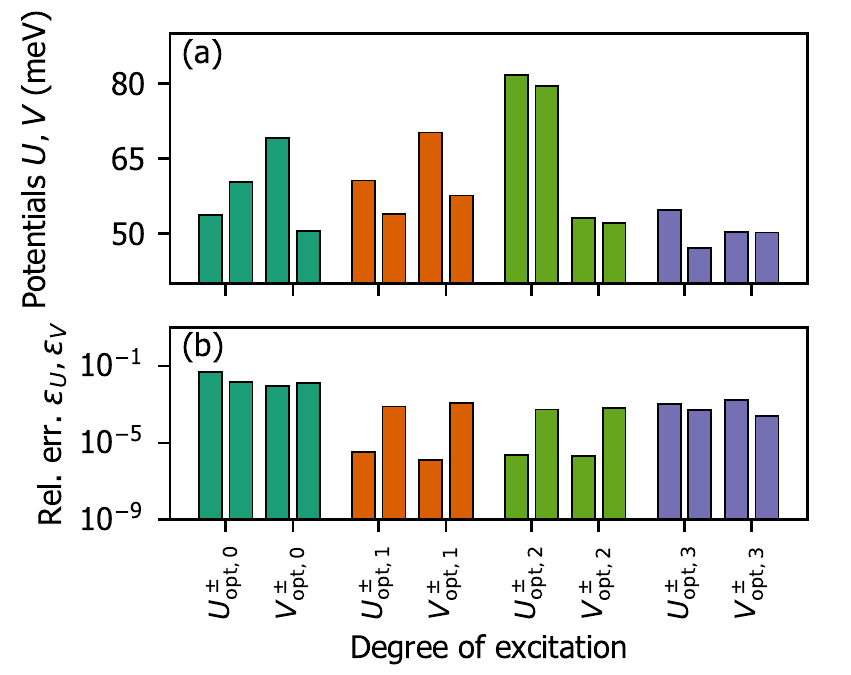}
    \caption{Fitted values (a) and associated statistical errors (b) of confining $U$ and gapping  $V$ potentials for QD1 as a result of the application of the HRS (CRS-LM) optimization routine.  $Q_{\rm opt,i}^\pm$ denotes the potential $Q\in\{U,V\}$ of the increasing (+) and decreasing (-) spectral line from the $i$-th spectral couple consisting of degenerated states at $B=0$~T.}   
    \label{fig:UV_values_errors_CG2}
\end{figure}

\begin{figure}[!tb]
    \centering
    \includegraphics{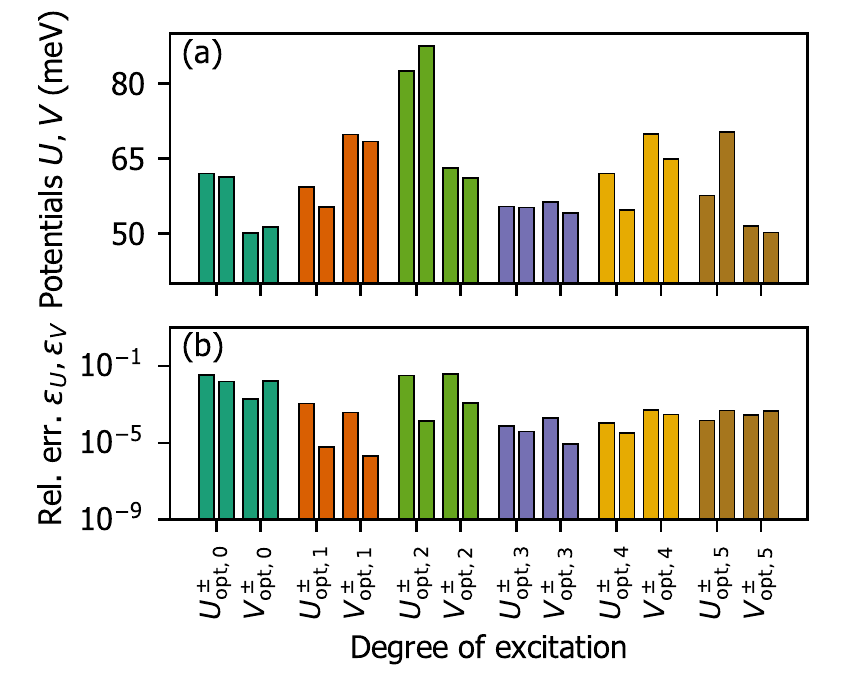}
    \caption{Fitted values (a) and associated statistical errors (b) of confining $U$ and gapping  $V$ potentials for QD2 as a result of the application of the HRS (CRS-LM) optimization routine.  $Q_{\rm opt,i}^\pm$ denotes the potential $Q\in\{U,V\}$ of the increasing (+) and decreasing (-) spectral line from the $i$-th spectral couple consisting of degenerated states at $B=0$~T.}   
    \label{fig:UV_values_errors_CG3}
\end{figure}

In the main text, we showed the final fitted energy states for all three quantum dot systems in Fig.~\ref{fig:experiments_all} and optimization details from continuous parameters search for the case of QD3 in Fig.~\ref{fig:UV_values_errors_CG9}. We present the optimization results analogous to those shown in Fig.~\ref{fig:UV_values_errors_CG9} for systems QD1 and QD2 in Figs.~\ref{fig:UV_values_errors_CG2}~and~\ref{fig:UV_values_errors_CG3}, respectively.  In these, we show the resulting potential values including their statistical errors of the inferred, continuous parameters $U$ and $V$. By and large, we reach a comparable precision in all cases.
Moreover, the recovered quantum numbers are in good agreement for all three dot systems and potential values remain within the expected value $\approx$50-80 meV. 
For completeness, we present all fitting parameters of the three quantum dots in Tab.~\ref{tab:U_V_values_and_errors}. 

\begin{table}[!ht]
\setlength{\tabcolsep}{4pt}
    \footnotesize	
    \centering
    \begin{tabular}{p{0.7cm}| r r c c c}
    & $m_{\rm opt}^\pm$ & $\tau_{\rm opt}^\pm$ & $U_{\rm opt}^\pm\, (\rm meV)$ & $V_{\rm opt}^\pm\, (\rm meV)$ & optimizer\\[0.1cm]
    \hline
    \hline
        \multirow{8}{*}{QD1 }& 0 & -1 &  53.7  $\pm$ 2.7e-00 & 69.1 $\pm$ 6.5e-01 & CRS-LM\\
        & 0 & 1 &  60.3  $\pm$ 8.7e-01 & 50.6 $\pm$ 6.8e-01 & CRS-LM\\
        & -2 & -1 &  60.6  $\pm$ 2.0e-04 & 70.3 $\pm$ 9.4e-05 & HRS\\
        & 2 & 1 &  53.9  $\pm$ 4.3e-02 & 57.6 $\pm$ 6.8e-02 & HRS\\
        & 1 & -1 &  81.7  $\pm$ 1.9e-04 & 53.1 $\pm$ 1.1e-04 & HRS\\
        & -1 & 1 &  79.5  $\pm$ 4.3e-02 & 52.0 $\pm$ 3.4e-02 & HRS\\
        & -1 & -1 &  54.6  $\pm$ 5.7e-02 & 50.2 $\pm$ 8.7e-02 & HRS\\
        & 1 & 1 &  47.1  $\pm$ 2.5e-02 & 50.2 $\pm$ 1.2e-02 & HRS\\
        \hline
        \multirow{12}{*}{QD2} & 0 & -1 &  62.0  $\pm$ 2.1e-00 & 50.1 $\pm$ 9.9e-02 & CRS-LM\\
        & 0 & 1 &  61.3  $\pm$ 9.6e-01 & 51.3 $\pm$ 8.9e-01 & CRS-LM\\
        & -2 & -1 &  59.4  $\pm$ 6.7e-02 & 69.8 $\pm$ 2.7e-02 & HRS\\
        & 2 & 1 &  55.3  $\pm$ 3.4e-04 & 68.4 $\pm$ 1.4e-04 & HRS\\
        & 1 & -1 &  82.5  $\pm$ 2.7e-00 & 63.2 $\pm$ 2.4e-00 & HRS\\
        & -1 & 1 &  87.6  $\pm$ 1.2e-02 & 61.2 $\pm$ 7.6e-02 & HRS\\
        & -1 & -1 &  55.4  $\pm$ 4.3e-03 & 56.2 $\pm$ 1.1e-02 & HRS\\
        & 1 & 1 &  55.2  $\pm$ 2.2e-03 & 54.2 $\pm$ 4.9e-04 & HRS\\
        & -3 & -1 &  62.0  $\pm$ 6.6e-03 & 70.0 $\pm$ 3.6e-02 & HRS\\
        & 3 & 1 &  54.7  $\pm$ 1.9e-03 & 64.9 $\pm$ 2.0e-02 & HRS\\
        & -2 & 1 &  57.6  $\pm$ 8.5e-03 & 51.5 $\pm$ 1.5e-02 & HRS\\
        & 2 & -1 &  70.3  $\pm$ 3.4e-02 & 50.2 $\pm$ 2.4e-02 & HRS\\
        \hline
        \multirow{10}{*}{QD3} & 0 & -1 &  60.1  $\pm$ 2.4e-00 & 50.1 $\pm$ 1.4e-01 & CRS-LM\\
        & 0 & 1 &  61.3  $\pm$ 1.3e-00 & 51.6 $\pm$ 1.6e-00 & CRS-LM\\
        & -2 & -1 &  68.4  $\pm$ 1.7e-04 & 50.0 $\pm$ 7.7e-05 & HRS\\
        & 2 & 1 &  68.7  $\pm$ 2.9e-04 & 50.2 $\pm$ 1.2e-04 & HRS\\
        & 1 & -1 &  75.0  $\pm$ 1.5e-02 & 50.5 $\pm$ 1.9e-02 & HRS\\
        & -1 & 1 &  79.9  $\pm$ 2.3e-02 & 52.3 $\pm$ 2.5e-01 & HRS\\
        & -1 & -1 &  59.2  $\pm$ 1.5e-01 & 50.3 $\pm$ 2.0e-01 & HRS\\
        & 1 & 1 &  60.2  $\pm$ 1.0e-04 & 50.2 $\pm$ 4.4e-05 & HRS\\
        & -3 & -1 &  54.1  $\pm$ 2.6e-05 & 61.8 $\pm$ 1.9e-04 & HRS\\
        & 3 & 1 &  56.3  $\pm$ 1.0e-06 & 50.0 $\pm$ 1.0e-06 & HRS\\
    \end{tabular}
    \caption{Confining and gapping potentials, $U$ and $V$, of single-particle quantum states distinguished by angular momenta $m$ and valley numbers $\tau$ for three different quantum dots, QD1-3, as provided by the HRS and the CRS-LM (for cases with $m=0$) algorithms. The given statistical errors are  computed based on repeated, independent optimization runs.}
    \label{tab:U_V_values_and_errors}
\end{table}

\section{Ground state quantum numbers with Deep Neural Networks (DNNs)}\label{app:nn_discrete}

\subsection{Motivation}
In Sec.~\ref{sec:results} of the main text, we have described the procedure to obtain the Hamiltonian parameters for the lowest-lying energy-lines couple $(m^\pm_{\rm opt,0},\tau^\pm_{\rm opt,0})$. Due to the unknown energy scale shift between experimental and model-based data, we had to rely on the plain CRS-LM algorithm. We note that fixing this energy scale also plays a crucial role as an initial step when fitting the excited states. In this section, we introduce an alternative approach to infer the discrete quantum numbers of the lowest-lying energy-lines couple based on a supervised neural network classifier for varying continuous parameters $U$ and $V$. Since the energies are not directly accessible, we explore the corresponding gradients and correlations thereof as input features.

\subsection{Data set creation}
\label{subsec:Datasetcreation}

To create the training data set, we use the bilayer graphene quantum dot model, introduced in Sec.~\ref{sec:bi}, that allows us to compute energy lines $E(B)$. Based on the physical assumption that the lowest energy couple will not possess a very high angular momentum, we limit ourselves to the five smallest momenta values $m \in \{-2,-1,0,1,2\}$. In a similar spirit, we use valley numbers $\tau\pm 1$ and we consider potentials $U,V$ in the range of $\{50,...,70\}$~meV~\cite{eich_2018}. We discretize these ranges by 90 steps, generate $91^2$ determinant maps for each of the 10 classes defined by all combinations of $m$ and $\tau$, and extract all spectral lines from the generated maps. As only the gradient information can be used, we transform the resulting lines into their gradients approximated as
\begin{align*}
    E(B_i) \rightarrow E(B_{i+1}) - E(B_i),
\end{align*}
where $B_i$ is $i$-th grid point along the $B$-axis. Here, we omit $\Delta B$ as our grid is evenly spaced. 

As we have discussed in Sec.~\ref{sec:results}, we fit the energy states in couples composed of (time-reversal) symmetric states, i.e., we require $m^+_i = - m^-_i$ and $\tau^+_i  =-\tau^-_i$ for the $i$-th state couple. Therefore, a single data point of our data set generally contains energy gradients from both the decreasing $\nabla E^-$ and the increasing $\nabla E^+$ line composing a symmetric couple. Additionally, we include the cross-correlation of the respective lines 
\begin{equation*}
        {\rm Corr}_{\nabla E^-,\nabla E^+}(k) = \sum_{l=1}^L \nabla E^-_l \nabla E^+_{l+k},
\end{equation*}
anticipating that the correlation between such two lines represent a powerful feature, which the network should consider to make predictions that generalize well to unseen data. Here, $L$ is the length of the gradient vectors and $k$ runs from $-L+1$ to $L-1$. In total, we have a single data point $x_i$ of our data set composed from three vectors:
\begin{equation*}
    x_i = (\nabla E^-,\nabla E^+,{\rm Corr_{\nabla E^-,\nabla E^+}}).
\end{equation*}
Finally, our training data set consists of $\sim 145000$ vectors of above structure representing the lowest-lying energy couples for ten different discrete quantum number candidates.

\subsection{Architecture, optimization, and results}

We employ a simple feed-forward network with two hidden layers. The hyperparameters specifying the DNN architecture are summarized in Tab.~\ref{tab:MLP_params}.

\begin{figure}
    \centering
    \includegraphics{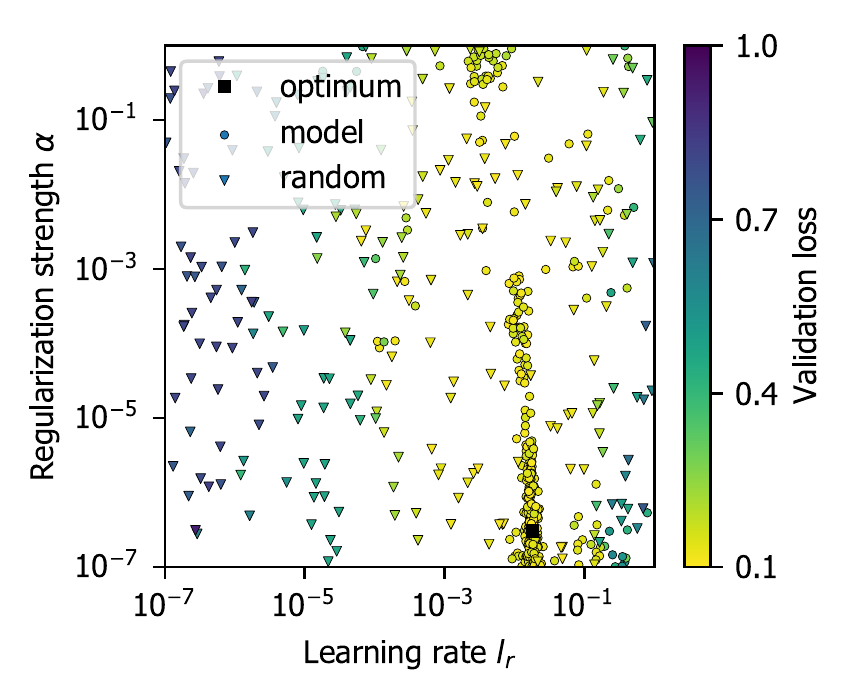}
    \caption{Optimization of the learning rate $l_r$ and regularization strength $\alpha$ using BOHB. Random (triangles), model picks (circles) as well as the optimal found configuration (black square) are displayed.}
    \label{fig:bohb}
\end{figure}

Most of the network hyperparameters were defined empirically. However, to determine the learning rate $l_r$ and the regularization strength $\alpha$, we used the Bayesian Optimization and Hyperband (BOHB) method~\cite{falkner2018bohb}. BOHB employs the successive halving method and improves random configuration picks with model-based ones resulting from the Bayesian model being fitted on the run.
The optimization landscape with all random and model picks are shown in Fig.~\ref{fig:bohb}. We split the data set into a training (80\%),  a validation (10\%), and a test set (10\%). Moreover, we use the classification accuracy $p_{\rm valid/ test}$:
\begin{equation*}
    p_{\rm valid/ test} = \frac{n_{\rm valid/test}(y_{\rm pred}= y_{\rm true})}{n_{\rm valid/test}},
\end{equation*} 
as the validation or test loss, respectively.

Within this setup, we trained 10 randomly initialized DNNs independently. In all cases, we reached a test accuracy of about 92.0\%-92.5\%.

\begin{figure}
    \centering
    \includegraphics{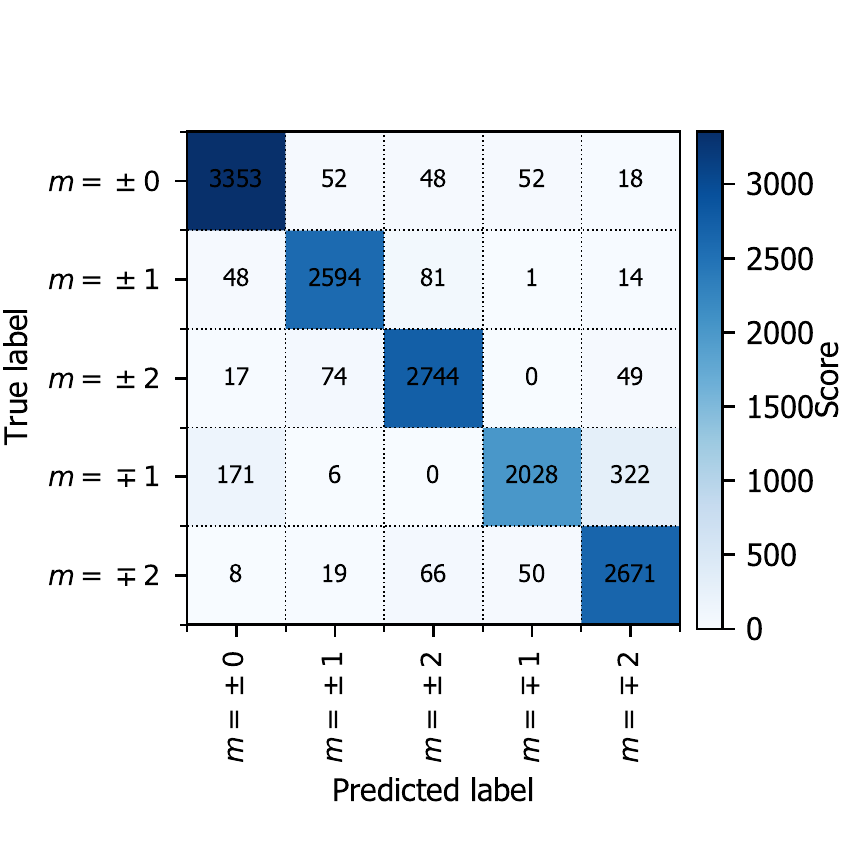}
    \caption{Confusion matrix for the testing of the DNN trained to predict the discrete quantum numbers of the lowest-lying energy state couple. Overall, the DNN reaches an accuracy of about 92.5\%.}
    \label{fig:conf_matrix}
\end{figure}

The experimental data (a single lowest-lying energy couple per quantum dot) was also pre-processed according to the steps outlined in~Sec.~\ref{subsec:Datasetcreation}. For all three dots, we obtain the prediction ($m^+=0, \tau^+=1$, $m^-=0, \tau^-=-1$) across all 10 trained networks with $>99\%$ confidence, which agrees with the result in the main text. For completeness, we show the test confusion matrix for one of the trained networks in Fig.~\ref{fig:conf_matrix}.  
In summary, we thus have an additional, independent approach to predict the discrete quantum numbers of the lowest-lying energy couple that is consistent across all three quantum dot measurements.

\begin{table}[!htb]
    \centering
    \begin{tabular}{|l |l|}
    \hline
    Input data  & shape $(144856, 173)$ \\
    \hline
    \multirow{3}{*}{$1^{\text{st}}$ dense} & type: linear \\
       & size: 800 \\
       & activation: ReLU\\
    \hline   
    \multirow{3}{*}{$2^{\text{nd}}$ dense} & type: linear \\
       & size: 800 \\
       & activation: ReLU\\
    \hline   
    \multirow{3}{*}{Output layer} & type: linear \\
       & size: 5 \\
       & activation: Softmax\\
    \hline
    Early stopping & 30 epochs\\
    \hline
    \multirow{2}{*}{Optimizer} & type: Adam\\
     & learning rate: 0.0182\\
    \hline 
    Loss & Cross Entropy  \\
    \hline
    Epochs & 60\\
    \hline
    Batch size & 200\\
    \hline
    Regularization coef. & $3.08 \cdot 10^{-7}$\\
    \hline
    \end{tabular}

    \caption{Hyperparameters of the DNN to predict the discrete quantum numbers of the lowest-lying energy-state couple, based on the approach introduced in Appendix~\ref{app:nn_discrete}.}
    \label{tab:MLP_params}
\end{table}

\end{appendix}

\clearpage
\bibliography{references}

\begin{thebibliography}{49}%
\makeatletter
\providecommand \@ifxundefined [1]{%
 \@ifx{#1\undefined}
}%
\providecommand \@ifnum [1]{%
 \ifnum #1\expandafter \@firstoftwo
 \else \expandafter \@secondoftwo
 \fi
}%
\providecommand \@ifx [1]{%
 \ifx #1\expandafter \@firstoftwo
 \else \expandafter \@secondoftwo
 \fi
}%
\providecommand \natexlab [1]{#1}%
\providecommand \enquote  [1]{``#1''}%
\providecommand \bibnamefont  [1]{#1}%
\providecommand \bibfnamefont [1]{#1}%
\providecommand \citenamefont [1]{#1}%
\providecommand \href@noop [0]{\@secondoftwo}%
\providecommand \href [0]{\begingroup \@sanitize@url \@href}%
\providecommand \@href[1]{\@@startlink{#1}\@@href}%
\providecommand \@@href[1]{\endgroup#1\@@endlink}%
\providecommand \@sanitize@url [0]{\catcode `\\12\catcode `\$12\catcode
  `\&12\catcode `\#12\catcode `\^12\catcode `\_12\catcode `\%12\relax}%
\providecommand \@@startlink[1]{}%
\providecommand \@@endlink[0]{}%
\providecommand \url  [0]{\begingroup\@sanitize@url \@url }%
\providecommand \@url [1]{\endgroup\@href {#1}{\urlprefix }}%
\providecommand \urlprefix  [0]{URL }%
\providecommand \Eprint [0]{\href }%
\providecommand \doibase [0]{http://dx.doi.org/}%
\providecommand \selectlanguage [0]{\@gobble}%
\providecommand \bibinfo  [0]{\@secondoftwo}%
\providecommand \bibfield  [0]{\@secondoftwo}%
\providecommand \translation [1]{[#1]}%
\providecommand \BibitemOpen [0]{}%
\providecommand \bibitemStop [0]{}%
\providecommand \bibitemNoStop [0]{.\EOS\space}%
\providecommand \EOS [0]{\spacefactor3000\relax}%
\providecommand \BibitemShut  [1]{\csname bibitem#1\endcsname}%
\let\auto@bib@innerbib\@empty
\bibitem [{\citenamefont {Novoselov}\ \emph {et~al.}(2016)\citenamefont
  {Novoselov}, \citenamefont {Mishchenko}, \citenamefont {Carvalho},\ and\
  \citenamefont {Neto}}]{novoselov20162d}%
  \BibitemOpen
  \bibfield  {author} {\bibinfo {author} {\bibfnamefont {K.~S.}\ \bibnamefont
  {Novoselov}}, \bibinfo {author} {\bibfnamefont {A.}~\bibnamefont
  {Mishchenko}}, \bibinfo {author} {\bibfnamefont {A.}~\bibnamefont
  {Carvalho}}, \ and\ \bibinfo {author} {\bibfnamefont {A.~H.~Castro}\
  \bibnamefont {Neto}},\ }\bibfield  {title} {\enquote {\bibinfo {title} {2{D}
  materials and van der {Waals} heterostructures},}\ }\href {\doibase
  10.1126/science.aac9439} {\bibfield  {journal} {\bibinfo  {journal}
  {Science}\ }\textbf {\bibinfo {volume} {353}} (\bibinfo {year} {2016}),\
  10.1126/science.aac9439}\BibitemShut {NoStop}%
\bibitem [{\citenamefont {Liu}\ \emph {et~al.}(2016)\citenamefont {Liu},
  \citenamefont {Weiss}, \citenamefont {Duan}, \citenamefont {Cheng},
  \citenamefont {Huang},\ and\ \citenamefont {Duan}}]{liu2016van}%
  \BibitemOpen
  \bibfield  {author} {\bibinfo {author} {\bibfnamefont {Y.}~\bibnamefont
  {Liu}}, \bibinfo {author} {\bibfnamefont {N.~O.}\ \bibnamefont {Weiss}},
  \bibinfo {author} {\bibfnamefont {X.}~\bibnamefont {Duan}}, \bibinfo {author}
  {\bibfnamefont {H.}~\bibnamefont {Cheng}}, \bibinfo {author} {\bibfnamefont
  {Y.}~\bibnamefont {Huang}}, \ and\ \bibinfo {author} {\bibfnamefont
  {X.}~\bibnamefont {Duan}},\ }\bibfield  {title} {\enquote {\bibinfo {title}
  {Van der {Waals} heterostructures and devices},}\ }\href {\doibase
  10.1038/natrevmats.2016.42} {\bibfield  {journal} {\bibinfo  {journal}
  {Nature Reviews Materials}\ }\textbf {\bibinfo {volume} {1}},\ \bibinfo
  {pages} {1--17} (\bibinfo {year} {2016})}\BibitemShut {NoStop}%
\bibitem [{\citenamefont {Liu}\ and\ \citenamefont {Hersam}(2019)}]{liu20192d}%
  \BibitemOpen
  \bibfield  {author} {\bibinfo {author} {\bibfnamefont {X.}~\bibnamefont
  {Liu}}\ and\ \bibinfo {author} {\bibfnamefont {M.~C.}\ \bibnamefont
  {Hersam}},\ }\bibfield  {title} {\enquote {\bibinfo {title} {2{D} materials
  for quantum information science},}\ }\href {\doibase
  10.1038/s41578-019-0136-x} {\bibfield  {journal} {\bibinfo  {journal} {Nature
  Reviews Materials}\ }\textbf {\bibinfo {volume} {4}},\ \bibinfo {pages}
  {669--684} (\bibinfo {year} {2019})}\BibitemShut {NoStop}%
\bibitem [{\citenamefont {Kennes}\ \emph {et~al.}(2021)\citenamefont {Kennes},
  \citenamefont {Claassen}, \citenamefont {Xian}, \citenamefont {Georges},
  \citenamefont {Millis}, \citenamefont {Hone}, \citenamefont {Dean},
  \citenamefont {Basov}, \citenamefont {Pasupathy},\ and\ \citenamefont
  {Rubio}}]{kennes2021moire}%
  \BibitemOpen
  \bibfield  {author} {\bibinfo {author} {\bibfnamefont {D.~M.}\ \bibnamefont
  {Kennes}}, \bibinfo {author} {\bibfnamefont {M.}~\bibnamefont {Claassen}},
  \bibinfo {author} {\bibfnamefont {L.}~\bibnamefont {Xian}}, \bibinfo {author}
  {\bibfnamefont {A.}~\bibnamefont {Georges}}, \bibinfo {author} {\bibfnamefont
  {A.~J.}\ \bibnamefont {Millis}}, \bibinfo {author} {\bibfnamefont
  {J.}~\bibnamefont {Hone}}, \bibinfo {author} {\bibfnamefont {C.~R.}\
  \bibnamefont {Dean}}, \bibinfo {author} {\bibfnamefont {D.N.}\ \bibnamefont
  {Basov}}, \bibinfo {author} {\bibfnamefont {A.~N.}\ \bibnamefont
  {Pasupathy}}, \ and\ \bibinfo {author} {\bibfnamefont {A.}~\bibnamefont
  {Rubio}},\ }\bibfield  {title} {\enquote {\bibinfo {title} {Moir{\'e}
  heterostructures as a condensed-matter quantum simulator},}\ }\href {\doibase
  10.1038/s41567-020-01154-3} {\bibfield  {journal} {\bibinfo  {journal}
  {Nature Physics}\ }\textbf {\bibinfo {volume} {17}},\ \bibinfo {pages} {1--9}
  (\bibinfo {year} {2021})}\BibitemShut {NoStop}%
\bibitem [{\citenamefont {Ohta}\ \emph
  {et~al.}(2006{\natexlab{a}})\citenamefont {Ohta}, \citenamefont {Bostwick},
  \citenamefont {Seyller}, \citenamefont {Horn},\ and\ \citenamefont
  {Rotenberg}}]{ohta2006controlling}%
  \BibitemOpen
  \bibfield  {author} {\bibinfo {author} {\bibfnamefont {T.}~\bibnamefont
  {Ohta}}, \bibinfo {author} {\bibfnamefont {A.}~\bibnamefont {Bostwick}},
  \bibinfo {author} {\bibfnamefont {T.}~\bibnamefont {Seyller}}, \bibinfo
  {author} {\bibfnamefont {K.}~\bibnamefont {Horn}}, \ and\ \bibinfo {author}
  {\bibfnamefont {E.}~\bibnamefont {Rotenberg}},\ }\bibfield  {title} {\enquote
  {\bibinfo {title} {Controlling the electronic structure of bilayer
  graphene},}\ }\href {\doibase 10.1126/science.1130681} {\bibfield  {journal}
  {\bibinfo  {journal} {Science}\ }\textbf {\bibinfo {volume} {313}},\ \bibinfo
  {pages} {951--954} (\bibinfo {year} {2006}{\natexlab{a}})}\BibitemShut
  {NoStop}%
\bibitem [{\citenamefont {McCann}\ and\ \citenamefont
  {Koshino}(2013)}]{mccann2013electronic}%
  \BibitemOpen
  \bibfield  {author} {\bibinfo {author} {\bibfnamefont {E.}~\bibnamefont
  {McCann}}\ and\ \bibinfo {author} {\bibfnamefont {M.}~\bibnamefont
  {Koshino}},\ }\bibfield  {title} {\enquote {\bibinfo {title} {The electronic
  properties of bilayer graphene},}\ }\href {\doibase
  10.1088/0034-4885/76/5/056503} {\bibfield  {journal} {\bibinfo  {journal}
  {Reports on Progress in Physics}\ }\textbf {\bibinfo {volume} {76}},\
  \bibinfo {pages} {056503} (\bibinfo {year} {2013})}\BibitemShut {NoStop}%
\bibitem [{\citenamefont {Bucko}\ and\ \citenamefont
  {Herman}(2021)}]{Bucko_2021}%
  \BibitemOpen
  \bibfield  {author} {\bibinfo {author} {\bibfnamefont {J.}~\bibnamefont
  {Bucko}}\ and\ \bibinfo {author} {\bibfnamefont {F.}~\bibnamefont {Herman}},\
  }\bibfield  {title} {\enquote {\bibinfo {title} {Large twisting angles in
  bilayer graphene moiré quantum dot structures},}\ }\href {\doibase
  https://doi.org/10.1103/PhysRevB.103.075116} {\bibfield  {journal} {\bibinfo
  {journal} {Physical Review B}\ }\textbf {\bibinfo {volume} {103}},\ \bibinfo
  {pages} {075116} (\bibinfo {year} {2021})}\BibitemShut {NoStop}%
\bibitem [{\citenamefont {Cao}\ \emph {et~al.}(2018)\citenamefont {Cao},
  \citenamefont {Fatemi}, \citenamefont {Fang}, \citenamefont {Watanabe},
  \citenamefont {Taniguchi}, \citenamefont {Kaxiras},\ and\ \citenamefont
  {Jarillo-Herrero}}]{cao2018unconventional}%
  \BibitemOpen
  \bibfield  {author} {\bibinfo {author} {\bibfnamefont {Y.}~\bibnamefont
  {Cao}}, \bibinfo {author} {\bibfnamefont {V.}~\bibnamefont {Fatemi}},
  \bibinfo {author} {\bibfnamefont {S.}~\bibnamefont {Fang}}, \bibinfo {author}
  {\bibfnamefont {K.}~\bibnamefont {Watanabe}}, \bibinfo {author}
  {\bibfnamefont {T.}~\bibnamefont {Taniguchi}}, \bibinfo {author}
  {\bibfnamefont {E.}~\bibnamefont {Kaxiras}}, \ and\ \bibinfo {author}
  {\bibfnamefont {P.}~\bibnamefont {Jarillo-Herrero}},\ }\bibfield  {title}
  {\enquote {\bibinfo {title} {Unconventional superconductivity in magic-angle
  graphene superlattices},}\ }\href {https://doi.org/10.1038/nature26160}
  {\bibfield  {journal} {\bibinfo  {journal} {Nature}\ }\textbf {\bibinfo
  {volume} {556}},\ \bibinfo {pages} {43--50} (\bibinfo {year}
  {2018})}\BibitemShut {NoStop}%
\bibitem [{\citenamefont {Wolf}\ \emph {et~al.}(2019)\citenamefont {Wolf},
  \citenamefont {Lado}, \citenamefont {Blatter},\ and\ \citenamefont
  {Zilberberg}}]{PhysRevLett.123.096802}%
  \BibitemOpen
  \bibfield  {author} {\bibinfo {author} {\bibfnamefont {T.~M.~R.}\
  \bibnamefont {Wolf}}, \bibinfo {author} {\bibfnamefont {J.~L.}\ \bibnamefont
  {Lado}}, \bibinfo {author} {\bibfnamefont {G.}~\bibnamefont {Blatter}}, \
  and\ \bibinfo {author} {\bibfnamefont {O.}~\bibnamefont {Zilberberg}},\
  }\bibfield  {title} {\enquote {\bibinfo {title} {Electrically tunable flat
  bands and magnetism in twisted bilayer graphene},}\ }\href {\doibase
  10.1103/PhysRevLett.123.096802} {\bibfield  {journal} {\bibinfo  {journal}
  {Phys. Rev. Lett.}\ }\textbf {\bibinfo {volume} {123}},\ \bibinfo {pages}
  {096802} (\bibinfo {year} {2019})}\BibitemShut {NoStop}%
\bibitem [{\citenamefont {Eich}\ \emph
  {et~al.}(2018{\natexlab{a}})\citenamefont {Eich}, \citenamefont {Pisoni},
  \citenamefont {Overweg}, \citenamefont {Kurzmann}, \citenamefont {Lee},
  \citenamefont {Rickhaus}, \citenamefont {Ihn}, \citenamefont {Ensslin},
  \citenamefont {Herman}, \citenamefont {Sigrist}, \citenamefont {Watanabe},\
  and\ \citenamefont {Taniguchi}}]{eich_2018}%
  \BibitemOpen
  \bibfield  {author} {\bibinfo {author} {\bibfnamefont {M.}~\bibnamefont
  {Eich}}, \bibinfo {author} {\bibfnamefont {R.}~\bibnamefont {Pisoni}},
  \bibinfo {author} {\bibfnamefont {H.}~\bibnamefont {Overweg}}, \bibinfo
  {author} {\bibfnamefont {A.}~\bibnamefont {Kurzmann}}, \bibinfo {author}
  {\bibfnamefont {Y.}~\bibnamefont {Lee}}, \bibinfo {author} {\bibfnamefont
  {P.}~\bibnamefont {Rickhaus}}, \bibinfo {author} {\bibfnamefont
  {T.}~\bibnamefont {Ihn}}, \bibinfo {author} {\bibfnamefont {K.}~\bibnamefont
  {Ensslin}}, \bibinfo {author} {\bibfnamefont {F.}~\bibnamefont {Herman}},
  \bibinfo {author} {\bibfnamefont {M.}~\bibnamefont {Sigrist}}, \bibinfo
  {author} {\bibfnamefont {K.}~\bibnamefont {Watanabe}}, \ and\ \bibinfo
  {author} {\bibfnamefont {T.}~\bibnamefont {Taniguchi}},\ }\bibfield  {title}
  {\enquote {\bibinfo {title} {Spin and valley states in gate-defined bilayer
  graphene quantum dots},}\ }\href {https://doi.org/10.1103/PhysRevX.8.031023}
  {\bibfield  {journal} {\bibinfo  {journal} {Physical Review X}\ }\textbf
  {\bibinfo {volume} {8}},\ \bibinfo {pages} {031023} (\bibinfo {year}
  {2018}{\natexlab{a}})}\BibitemShut {NoStop}%
\bibitem [{\citenamefont {Hill}\ \emph {et~al.}(2015)\citenamefont {Hill},
  \citenamefont {Peretz}, \citenamefont {Hile}, \citenamefont {House},
  \citenamefont {Fuechsle}, \citenamefont {Rogge}, \citenamefont {Simmons},\
  and\ \citenamefont {Hollenberg}}]{hill2015surface}%
  \BibitemOpen
  \bibfield  {author} {\bibinfo {author} {\bibfnamefont {Ch.~D.}\ \bibnamefont
  {Hill}}, \bibinfo {author} {\bibfnamefont {E.}~\bibnamefont {Peretz}},
  \bibinfo {author} {\bibfnamefont {S.~J.}\ \bibnamefont {Hile}}, \bibinfo
  {author} {\bibfnamefont {M.~G.}\ \bibnamefont {House}}, \bibinfo {author}
  {\bibfnamefont {M.}~\bibnamefont {Fuechsle}}, \bibinfo {author}
  {\bibfnamefont {S.}~\bibnamefont {Rogge}}, \bibinfo {author} {\bibfnamefont
  {M.~Y.}\ \bibnamefont {Simmons}}, \ and\ \bibinfo {author} {\bibfnamefont
  {L.~C.L.}\ \bibnamefont {Hollenberg}},\ }\bibfield  {title} {\enquote
  {\bibinfo {title} {A surface code quantum computer in silicon},}\ }\href
  {\doibase 10.1126/sciadv.1500707} {\bibfield  {journal} {\bibinfo  {journal}
  {Science Advances}\ }\textbf {\bibinfo {volume} {1}},\ \bibinfo {pages}
  {e1500707} (\bibinfo {year} {2015})}\BibitemShut {NoStop}%
\bibitem [{\citenamefont {Watson}\ \emph {et~al.}(2018)\citenamefont {Watson},
  \citenamefont {Philips}, \citenamefont {Kawakami}, \citenamefont {Ward},
  \citenamefont {Scarlino}, \citenamefont {Veldhorst}, \citenamefont {Savage},
  \citenamefont {Lagally}, \citenamefont {Friesen}, \citenamefont {Coppersmith}
  \emph {et~al.}}]{watson2018programmable}%
  \BibitemOpen
  \bibfield  {author} {\bibinfo {author} {\bibfnamefont {T.~F.}\ \bibnamefont
  {Watson}}, \bibinfo {author} {\bibfnamefont {S.~G.~J.}\ \bibnamefont
  {Philips}}, \bibinfo {author} {\bibfnamefont {E.}~\bibnamefont {Kawakami}},
  \bibinfo {author} {\bibfnamefont {D.~R.}\ \bibnamefont {Ward}}, \bibinfo
  {author} {\bibfnamefont {P.}~\bibnamefont {Scarlino}}, \bibinfo {author}
  {\bibfnamefont {M.}~\bibnamefont {Veldhorst}}, \bibinfo {author}
  {\bibfnamefont {D.~E.}\ \bibnamefont {Savage}}, \bibinfo {author}
  {\bibfnamefont {M.~G.}\ \bibnamefont {Lagally}}, \bibinfo {author}
  {\bibfnamefont {M.}~\bibnamefont {Friesen}}, \bibinfo {author} {\bibfnamefont
  {S.~N.}\ \bibnamefont {Coppersmith}},  \emph {et~al.},\ }\bibfield  {title}
  {\enquote {\bibinfo {title} {A programmable two-qubit quantum processor in
  silicon},}\ }\href {https://doi.org/10.1038/nature25766} {\bibfield
  {journal} {\bibinfo  {journal} {Nature}\ }\textbf {\bibinfo {volume} {555}},\
  \bibinfo {pages} {633--637} (\bibinfo {year} {2018})}\BibitemShut {NoStop}%
\bibitem [{\citenamefont {Loss}\ and\ \citenamefont
  {DiVincenzo}(1998)}]{Loss_1998}%
  \BibitemOpen
  \bibfield  {author} {\bibinfo {author} {\bibfnamefont {D.}~\bibnamefont
  {Loss}}\ and\ \bibinfo {author} {\bibfnamefont {D.~P.}\ \bibnamefont
  {DiVincenzo}},\ }\bibfield  {title} {\enquote {\bibinfo {title} {Quantum
  computation with quantum dots},}\ }\href {\doibase
  https://doi.org/10.1103/PhysRevA.57.120} {\bibfield  {journal} {\bibinfo
  {journal} {Physical Review A}\ }\textbf {\bibinfo {volume} {57}},\ \bibinfo
  {pages} {120} (\bibinfo {year} {1998})}\BibitemShut {NoStop}%
\bibitem [{\citenamefont {Trauzettel}\ \emph {et~al.}(2007)\citenamefont
  {Trauzettel}, \citenamefont {Bulaev}, \citenamefont {Loss},\ and\
  \citenamefont {Burkard}}]{Trauzettel_2007}%
  \BibitemOpen
  \bibfield  {author} {\bibinfo {author} {\bibfnamefont {B.}~\bibnamefont
  {Trauzettel}}, \bibinfo {author} {\bibfnamefont {D.~V.}\ \bibnamefont
  {Bulaev}}, \bibinfo {author} {\bibfnamefont {D.}~\bibnamefont {Loss}}, \ and\
  \bibinfo {author} {\bibfnamefont {G.}~\bibnamefont {Burkard}},\ }\bibfield
  {title} {\enquote {\bibinfo {title} {Spin qubits in graphene quantum dots},}\
  }\href {\doibase 10.1038/nphys544} {\bibfield  {journal} {\bibinfo  {journal}
  {Nature}\ }\textbf {\bibinfo {volume} {3}},\ \bibinfo {pages} {192 -- 196}
  (\bibinfo {year} {2007})}\BibitemShut {NoStop}%
\bibitem [{\citenamefont {Eich}\ \emph
  {et~al.}(2018{\natexlab{b}})\citenamefont {Eich}, \citenamefont {Pisoni},
  \citenamefont {Pally}, \citenamefont {Overweg}, \citenamefont {Kurzmann},
  \citenamefont {Lee}, \citenamefont {Rickhaus}, \citenamefont {Watanabe},
  \citenamefont {Taniguchi}, \citenamefont {Ensslin} \emph
  {et~al.}}]{eich2018coupled}%
  \BibitemOpen
  \bibfield  {author} {\bibinfo {author} {\bibfnamefont {M.}~\bibnamefont
  {Eich}}, \bibinfo {author} {\bibfnamefont {R.}~\bibnamefont {Pisoni}},
  \bibinfo {author} {\bibfnamefont {A.}~\bibnamefont {Pally}}, \bibinfo
  {author} {\bibfnamefont {H.}~\bibnamefont {Overweg}}, \bibinfo {author}
  {\bibfnamefont {A.}~\bibnamefont {Kurzmann}}, \bibinfo {author}
  {\bibfnamefont {Y.}~\bibnamefont {Lee}}, \bibinfo {author} {\bibfnamefont
  {P.}~\bibnamefont {Rickhaus}}, \bibinfo {author} {\bibfnamefont
  {K.}~\bibnamefont {Watanabe}}, \bibinfo {author} {\bibfnamefont
  {T.}~\bibnamefont {Taniguchi}}, \bibinfo {author} {\bibfnamefont
  {K.}~\bibnamefont {Ensslin}},  \emph {et~al.},\ }\bibfield  {title} {\enquote
  {\bibinfo {title} {Coupled quantum dots in bilayer graphene},}\ }\href
  {https://doi.org/10.1021/acs.nanolett.8b01859} {\bibfield  {journal}
  {\bibinfo  {journal} {Nano letters}\ }\textbf {\bibinfo {volume} {18}},\
  \bibinfo {pages} {5042--5048} (\bibinfo {year}
  {2018}{\natexlab{b}})}\BibitemShut {NoStop}%
\bibitem [{\citenamefont {Kurzmann}\ \emph {et~al.}(2019)\citenamefont
  {Kurzmann}, \citenamefont {Eich}, \citenamefont {Overweg}, \citenamefont
  {Mangold}, \citenamefont {Herman}, \citenamefont {Rickhaus}, \citenamefont
  {Pisoni}, \citenamefont {Lee}, \citenamefont {Garreis}, \citenamefont {Tong}
  \emph {et~al.}}]{kurzmann2019excited}%
  \BibitemOpen
  \bibfield  {author} {\bibinfo {author} {\bibfnamefont {A.}~\bibnamefont
  {Kurzmann}}, \bibinfo {author} {\bibfnamefont {M.}~\bibnamefont {Eich}},
  \bibinfo {author} {\bibfnamefont {H.}~\bibnamefont {Overweg}}, \bibinfo
  {author} {\bibfnamefont {M.}~\bibnamefont {Mangold}}, \bibinfo {author}
  {\bibfnamefont {F.}~\bibnamefont {Herman}}, \bibinfo {author} {\bibfnamefont
  {P.}~\bibnamefont {Rickhaus}}, \bibinfo {author} {\bibfnamefont
  {R.}~\bibnamefont {Pisoni}}, \bibinfo {author} {\bibfnamefont
  {Y.}~\bibnamefont {Lee}}, \bibinfo {author} {\bibfnamefont {R.}~\bibnamefont
  {Garreis}}, \bibinfo {author} {\bibfnamefont {Ch.}\ \bibnamefont {Tong}},
  \emph {et~al.},\ }\bibfield  {title} {\enquote {\bibinfo {title} {Excited
  states in bilayer graphene quantum dots},}\ }\href
  {https://doi.org/10.1103/PhysRevLett.123.026803} {\bibfield  {journal}
  {\bibinfo  {journal} {Physical Review Letters}\ }\textbf {\bibinfo {volume}
  {123}},\ \bibinfo {pages} {026803} (\bibinfo {year} {2019})}\BibitemShut
  {NoStop}%
\bibitem [{\citenamefont {Greplova}\ \emph {et~al.}(2020)\citenamefont
  {Greplova}, \citenamefont {Gold}, \citenamefont {Kratochwil}, \citenamefont
  {Davatz}, \citenamefont {Pisoni}, \citenamefont {Kurzmann}, \citenamefont
  {Rickhaus}, \citenamefont {Fischer}, \citenamefont {Ihn},\ and\ \citenamefont
  {Huber}}]{Greplova_2020}%
  \BibitemOpen
  \bibfield  {author} {\bibinfo {author} {\bibfnamefont {E.}~\bibnamefont
  {Greplova}}, \bibinfo {author} {\bibfnamefont {C.}~\bibnamefont {Gold}},
  \bibinfo {author} {\bibfnamefont {B.}~\bibnamefont {Kratochwil}}, \bibinfo
  {author} {\bibfnamefont {T.}~\bibnamefont {Davatz}}, \bibinfo {author}
  {\bibfnamefont {R.}~\bibnamefont {Pisoni}}, \bibinfo {author} {\bibfnamefont
  {A.}~\bibnamefont {Kurzmann}}, \bibinfo {author} {\bibfnamefont
  {P.}~\bibnamefont {Rickhaus}}, \bibinfo {author} {\bibfnamefont {M.~H.}\
  \bibnamefont {Fischer}}, \bibinfo {author} {\bibfnamefont {T.}~\bibnamefont
  {Ihn}}, \ and\ \bibinfo {author} {\bibfnamefont {S.~D.}\ \bibnamefont
  {Huber}},\ }\bibfield  {title} {\enquote {\bibinfo {title} {Fully automated
  identification of two-dimensional material samples},}\ }\href
  {https://Journals.aps.org/prapplied/abstract/10.1103/PhysRevApplied.13.064017}
  {\bibfield  {journal} {\bibinfo  {journal} {Physical Review Applied}\
  }\textbf {\bibinfo {volume} {13}},\ \bibinfo {pages} {064017} (\bibinfo
  {year} {2020})}\BibitemShut {NoStop}%
\bibitem [{\citenamefont {Banszerus}\ \emph {et~al.}(2021)\citenamefont
  {Banszerus}, \citenamefont {M{\"o}ller}, \citenamefont {Icking},
  \citenamefont {Steiner}, \citenamefont {Neumaier}, \citenamefont {Otto},
  \citenamefont {Watanabe}, \citenamefont {Taniguchi}, \citenamefont {Volk},\
  and\ \citenamefont {Stampfer}}]{banszerus2020dispersive}%
  \BibitemOpen
  \bibfield  {author} {\bibinfo {author} {\bibfnamefont {L.}~\bibnamefont
  {Banszerus}}, \bibinfo {author} {\bibfnamefont {S.}~\bibnamefont
  {M{\"o}ller}}, \bibinfo {author} {\bibfnamefont {E.}~\bibnamefont {Icking}},
  \bibinfo {author} {\bibfnamefont {C.}~\bibnamefont {Steiner}}, \bibinfo
  {author} {\bibfnamefont {D.}~\bibnamefont {Neumaier}}, \bibinfo {author}
  {\bibfnamefont {M.}~\bibnamefont {Otto}}, \bibinfo {author} {\bibfnamefont
  {K.}~\bibnamefont {Watanabe}}, \bibinfo {author} {\bibfnamefont
  {T.}~\bibnamefont {Taniguchi}}, \bibinfo {author} {\bibfnamefont {Ch.}\
  \bibnamefont {Volk}}, \ and\ \bibinfo {author} {\bibfnamefont {Ch.}\
  \bibnamefont {Stampfer}},\ }\bibfield  {title} {\enquote {\bibinfo {title}
  {Dispersive sensing of charge states in a bilayer graphene quantum dot},}\
  }\href {https://doi.org/10.1063/5.0040234} {\bibfield  {journal} {\bibinfo
  {journal} {Applied Physics Letters}\ }\textbf {\bibinfo {volume} {118}},\
  \bibinfo {pages} {093104} (\bibinfo {year} {2021})}\BibitemShut {NoStop}%
\bibitem [{\citenamefont {Lyon}\ \emph {et~al.}(2017)\citenamefont {Lyon},
  \citenamefont {Sichau}, \citenamefont {Dorn}, \citenamefont {Centeno},
  \citenamefont {Pesquera}, \citenamefont {Zurutuza},\ and\ \citenamefont
  {Blick}}]{lyon2017probing}%
  \BibitemOpen
  \bibfield  {author} {\bibinfo {author} {\bibfnamefont {T.~J.}\ \bibnamefont
  {Lyon}}, \bibinfo {author} {\bibfnamefont {J.}~\bibnamefont {Sichau}},
  \bibinfo {author} {\bibfnamefont {A.}~\bibnamefont {Dorn}}, \bibinfo {author}
  {\bibfnamefont {A.}~\bibnamefont {Centeno}}, \bibinfo {author} {\bibfnamefont
  {A.}~\bibnamefont {Pesquera}}, \bibinfo {author} {\bibfnamefont
  {A.}~\bibnamefont {Zurutuza}}, \ and\ \bibinfo {author} {\bibfnamefont
  {R.~H.}\ \bibnamefont {Blick}},\ }\bibfield  {title} {\enquote {\bibinfo
  {title} {Probing electron spin resonance in monolayer graphene},}\ }\href
  {https://doi.org/10.1103/PhysRevLett.119.066802} {\bibfield  {journal}
  {\bibinfo  {journal} {Physical Review Letters}\ }\textbf {\bibinfo {volume}
  {119}},\ \bibinfo {pages} {066802} (\bibinfo {year} {2017})}\BibitemShut
  {NoStop}%
\bibitem [{\citenamefont {Sichau}\ \emph {et~al.}(2019)\citenamefont {Sichau},
  \citenamefont {Prada}, \citenamefont {Anlauf}, \citenamefont {Lyon},
  \citenamefont {Bosnjak}, \citenamefont {Tiemann},\ and\ \citenamefont
  {Blick}}]{sichau2019resonance}%
  \BibitemOpen
  \bibfield  {author} {\bibinfo {author} {\bibfnamefont {J.}~\bibnamefont
  {Sichau}}, \bibinfo {author} {\bibfnamefont {M.}~\bibnamefont {Prada}},
  \bibinfo {author} {\bibfnamefont {T.}~\bibnamefont {Anlauf}}, \bibinfo
  {author} {\bibfnamefont {T.J.}\ \bibnamefont {Lyon}}, \bibinfo {author}
  {\bibfnamefont {B.}~\bibnamefont {Bosnjak}}, \bibinfo {author} {\bibfnamefont
  {L.}~\bibnamefont {Tiemann}}, \ and\ \bibinfo {author} {\bibfnamefont {R.H.}\
  \bibnamefont {Blick}},\ }\bibfield  {title} {\enquote {\bibinfo {title}
  {Resonance microwave measurements of an intrinsic spin-orbit coupling gap in
  graphene: {A} possible indication of a topological state},}\ }\href
  {https://doi.org/10.1103/PhysRevLett.122.046403} {\bibfield  {journal}
  {\bibinfo  {journal} {Physical Review Letters}\ }\textbf {\bibinfo {volume}
  {122}},\ \bibinfo {pages} {046403} (\bibinfo {year} {2019})}\BibitemShut
  {NoStop}%
\bibitem [{\citenamefont {Min}\ \emph {et~al.}(2006)\citenamefont {Min},
  \citenamefont {Hill}, \citenamefont {Sinitsyn}, \citenamefont {Sahu},
  \citenamefont {Kleinman},\ and\ \citenamefont {MacDonald}}]{Min_2006}%
  \BibitemOpen
  \bibfield  {author} {\bibinfo {author} {\bibfnamefont {H.}~\bibnamefont
  {Min}}, \bibinfo {author} {\bibfnamefont {J.~E.}\ \bibnamefont {Hill}},
  \bibinfo {author} {\bibfnamefont {N.~A.}\ \bibnamefont {Sinitsyn}}, \bibinfo
  {author} {\bibfnamefont {B.~R.}\ \bibnamefont {Sahu}}, \bibinfo {author}
  {\bibfnamefont {L.}~\bibnamefont {Kleinman}}, \ and\ \bibinfo {author}
  {\bibfnamefont {A.~H.}\ \bibnamefont {MacDonald}},\ }\bibfield  {title}
  {\enquote {\bibinfo {title} {Intrinsic and rashba spin-orbit interactions in
  graphene sheets},}\ }\href {\doibase
  https://doi.org/10.1103/PhysRevB.74.165310} {\bibfield  {journal} {\bibinfo
  {journal} {Physical Review B}\ }\textbf {\bibinfo {volume} {74}},\ \bibinfo
  {pages} {165310} (\bibinfo {year} {2006})}\BibitemShut {NoStop}%
\bibitem [{\citenamefont {Huertas-Hernando}\ \emph {et~al.}(2006)\citenamefont
  {Huertas-Hernando}, \citenamefont {Guinea},\ and\ \citenamefont
  {Brataas}}]{Hernando_2006}%
  \BibitemOpen
  \bibfield  {author} {\bibinfo {author} {\bibfnamefont {D.}~\bibnamefont
  {Huertas-Hernando}}, \bibinfo {author} {\bibfnamefont {F.}~\bibnamefont
  {Guinea}}, \ and\ \bibinfo {author} {\bibfnamefont {A.}~\bibnamefont
  {Brataas}},\ }\bibfield  {title} {\enquote {\bibinfo {title} {Spin-orbit
  coupling in curved graphene, fullerenes, nanotubes, and nanotube caps},}\
  }\href {\doibase https://doi.org/10.1103/PhysRevB.74.155426} {\bibfield
  {journal} {\bibinfo  {journal} {Physical Review B}\ }\textbf {\bibinfo
  {volume} {74}},\ \bibinfo {pages} {155426} (\bibinfo {year}
  {2006})}\BibitemShut {NoStop}%
\bibitem [{\citenamefont {Ohta}\ \emph
  {et~al.}(2006{\natexlab{b}})\citenamefont {Ohta}, \citenamefont {Bostwick},
  \citenamefont {Seyller}, \citenamefont {Horn},\ and\ \citenamefont
  {Rotenberg}}]{Ohta_2006}%
  \BibitemOpen
  \bibfield  {author} {\bibinfo {author} {\bibfnamefont {T.}~\bibnamefont
  {Ohta}}, \bibinfo {author} {\bibfnamefont {A.}~\bibnamefont {Bostwick}},
  \bibinfo {author} {\bibfnamefont {T.}~\bibnamefont {Seyller}}, \bibinfo
  {author} {\bibfnamefont {K.}~\bibnamefont {Horn}}, \ and\ \bibinfo {author}
  {\bibfnamefont {E.}~\bibnamefont {Rotenberg}},\ }\bibfield  {title} {\enquote
  {\bibinfo {title} {Controlling the electronic structure of bilayer
  graphene},}\ }\href {https://science.sciencemag.org/content/313/5789/951}
  {\bibfield  {journal} {\bibinfo  {journal} {Science}\ }\textbf {\bibinfo
  {volume} {313}},\ \bibinfo {pages} {951 -- 954} (\bibinfo {year}
  {2006}{\natexlab{b}})}\BibitemShut {NoStop}%
\bibitem [{\citenamefont {Rickhaus}\ \emph {et~al.}(2019)\citenamefont
  {Rickhaus}, \citenamefont {Zheng}, \citenamefont {Lado}, \citenamefont {Lee},
  \citenamefont {Kurzmann}, \citenamefont {Eich}, \citenamefont {Pisoni},
  \citenamefont {Tong}, \citenamefont {Garreis}, \citenamefont {Gold} \emph
  {et~al.}}]{rickhaus2019gap}%
  \BibitemOpen
  \bibfield  {author} {\bibinfo {author} {\bibfnamefont {P.}~\bibnamefont
  {Rickhaus}}, \bibinfo {author} {\bibfnamefont {G.}~\bibnamefont {Zheng}},
  \bibinfo {author} {\bibfnamefont {J.~L.}\ \bibnamefont {Lado}}, \bibinfo
  {author} {\bibfnamefont {Y.}~\bibnamefont {Lee}}, \bibinfo {author}
  {\bibfnamefont {A.}~\bibnamefont {Kurzmann}}, \bibinfo {author}
  {\bibfnamefont {M.}~\bibnamefont {Eich}}, \bibinfo {author} {\bibfnamefont
  {R.}~\bibnamefont {Pisoni}}, \bibinfo {author} {\bibfnamefont {Ch.}\
  \bibnamefont {Tong}}, \bibinfo {author} {\bibfnamefont {R.}~\bibnamefont
  {Garreis}}, \bibinfo {author} {\bibfnamefont {C.}~\bibnamefont {Gold}},
  \emph {et~al.},\ }\bibfield  {title} {\enquote {\bibinfo {title} {Gap opening
  in twisted double bilayer graphene by crystal fields},}\ }\href
  {https://doi.org/10.1021/acs.nanolett.9b03660} {\bibfield  {journal}
  {\bibinfo  {journal} {Nano Letters}\ }\textbf {\bibinfo {volume} {19}},\
  \bibinfo {pages} {8821--8828} (\bibinfo {year} {2019})}\BibitemShut {NoStop}%
\bibitem [{\citenamefont {Zwanenburg}\ \emph {et~al.}(2013)\citenamefont
  {Zwanenburg}, \citenamefont {Dzurak}, \citenamefont {Morello}, \citenamefont
  {Simmons}, \citenamefont {Hollenberg}, \citenamefont {Klimeck}, \citenamefont
  {Rogge}, \citenamefont {Coppersmith},\ and\ \citenamefont
  {Eriksson}}]{zwanenburg2013silicon}%
  \BibitemOpen
  \bibfield  {author} {\bibinfo {author} {\bibfnamefont {F.~A}\ \bibnamefont
  {Zwanenburg}}, \bibinfo {author} {\bibfnamefont {A.~S.}\ \bibnamefont
  {Dzurak}}, \bibinfo {author} {\bibfnamefont {A.}~\bibnamefont {Morello}},
  \bibinfo {author} {\bibfnamefont {M.~Y.}\ \bibnamefont {Simmons}}, \bibinfo
  {author} {\bibfnamefont {L.~C.~L.}\ \bibnamefont {Hollenberg}}, \bibinfo
  {author} {\bibfnamefont {G.}~\bibnamefont {Klimeck}}, \bibinfo {author}
  {\bibfnamefont {S.}~\bibnamefont {Rogge}}, \bibinfo {author} {\bibfnamefont
  {S.~N.}\ \bibnamefont {Coppersmith}}, \ and\ \bibinfo {author} {\bibfnamefont
  {M.~A.}\ \bibnamefont {Eriksson}},\ }\bibfield  {title} {\enquote {\bibinfo
  {title} {Silicon quantum electronics},}\ }\href
  {https://doi.org/10.1103/RevModPhys.85.961} {\bibfield  {journal} {\bibinfo
  {journal} {Reviews of Modern Physics}\ }\textbf {\bibinfo {volume} {85}},\
  \bibinfo {pages} {961} (\bibinfo {year} {2013})}\BibitemShut {NoStop}%
\bibitem [{\citenamefont {Veldhorst}\ \emph {et~al.}(2014)\citenamefont
  {Veldhorst}, \citenamefont {Hwang}, \citenamefont {Yang}, \citenamefont
  {Leenstra}, \citenamefont {de~Ronde}, \citenamefont {Dehollain},
  \citenamefont {Muhonen}, \citenamefont {Hudson}, \citenamefont {Itoh},
  \citenamefont {Morello} \emph {et~al.}}]{veldhorst2014addressable}%
  \BibitemOpen
  \bibfield  {author} {\bibinfo {author} {\bibfnamefont {M.}~\bibnamefont
  {Veldhorst}}, \bibinfo {author} {\bibfnamefont {J.~C.~C.}\ \bibnamefont
  {Hwang}}, \bibinfo {author} {\bibfnamefont {C.~H.}\ \bibnamefont {Yang}},
  \bibinfo {author} {\bibfnamefont {A.~W.}\ \bibnamefont {Leenstra}}, \bibinfo
  {author} {\bibfnamefont {B.}~\bibnamefont {de~Ronde}}, \bibinfo {author}
  {\bibfnamefont {J.~P.}\ \bibnamefont {Dehollain}}, \bibinfo {author}
  {\bibfnamefont {J.~T.}\ \bibnamefont {Muhonen}}, \bibinfo {author}
  {\bibfnamefont {F.~E.}\ \bibnamefont {Hudson}}, \bibinfo {author}
  {\bibfnamefont {K.~M.}\ \bibnamefont {Itoh}}, \bibinfo {author}
  {\bibfnamefont {A.}~\bibnamefont {Morello}},  \emph {et~al.},\ }\bibfield
  {title} {\enquote {\bibinfo {title} {An addressable quantum dot qubit with
  fault-tolerant control-fidelity},}\ }\href
  {https://doi.org/10.1038/nnano.2014.216} {\bibfield  {journal} {\bibinfo
  {journal} {Nature Nanotechnology}\ }\textbf {\bibinfo {volume} {9}},\
  \bibinfo {pages} {981--985} (\bibinfo {year} {2014})}\BibitemShut {NoStop}%
\bibitem [{\citenamefont {Yoneda}\ \emph {et~al.}(2018)\citenamefont {Yoneda},
  \citenamefont {Takeda}, \citenamefont {Otsuka}, \citenamefont {Nakajima},
  \citenamefont {Delbecq}, \citenamefont {Allison}, \citenamefont {Honda},
  \citenamefont {Kodera}, \citenamefont {Oda}, \citenamefont {Hoshi} \emph
  {et~al.}}]{yoneda2018quantum}%
  \BibitemOpen
  \bibfield  {author} {\bibinfo {author} {\bibfnamefont {J.}~\bibnamefont
  {Yoneda}}, \bibinfo {author} {\bibfnamefont {K.}~\bibnamefont {Takeda}},
  \bibinfo {author} {\bibfnamefont {T.}~\bibnamefont {Otsuka}}, \bibinfo
  {author} {\bibfnamefont {T.}~\bibnamefont {Nakajima}}, \bibinfo {author}
  {\bibfnamefont {M.~R.}\ \bibnamefont {Delbecq}}, \bibinfo {author}
  {\bibfnamefont {G.}~\bibnamefont {Allison}}, \bibinfo {author} {\bibfnamefont
  {T.}~\bibnamefont {Honda}}, \bibinfo {author} {\bibfnamefont
  {T.}~\bibnamefont {Kodera}}, \bibinfo {author} {\bibfnamefont
  {S.}~\bibnamefont {Oda}}, \bibinfo {author} {\bibfnamefont {Y.}~\bibnamefont
  {Hoshi}},  \emph {et~al.},\ }\bibfield  {title} {\enquote {\bibinfo {title}
  {A quantum-dot spin qubit with coherence limited by charge noise and fidelity
  higher than 99.9\%},}\ }\href {https://doi.org/10.1038/s41565-017-0014-x}
  {\bibfield  {journal} {\bibinfo  {journal} {Nature Nanotechnology}\ }\textbf
  {\bibinfo {volume} {13}},\ \bibinfo {pages} {102--106} (\bibinfo {year}
  {2018})}\BibitemShut {NoStop}%
\bibitem [{\citenamefont {Landig}\ \emph {et~al.}(2018)\citenamefont {Landig},
  \citenamefont {Koski}, \citenamefont {Scarlino}, \citenamefont {Mendes},
  \citenamefont {Blais}, \citenamefont {Reichl}, \citenamefont {Wegscheider},
  \citenamefont {Wallraff}, \citenamefont {Ensslin},\ and\ \citenamefont
  {Ihn}}]{landig2018coherent}%
  \BibitemOpen
  \bibfield  {author} {\bibinfo {author} {\bibfnamefont {A.~J.}\ \bibnamefont
  {Landig}}, \bibinfo {author} {\bibfnamefont {J.~V.}\ \bibnamefont {Koski}},
  \bibinfo {author} {\bibfnamefont {P.}~\bibnamefont {Scarlino}}, \bibinfo
  {author} {\bibfnamefont {U.~C.}\ \bibnamefont {Mendes}}, \bibinfo {author}
  {\bibfnamefont {A.}~\bibnamefont {Blais}}, \bibinfo {author} {\bibfnamefont
  {Ch.}\ \bibnamefont {Reichl}}, \bibinfo {author} {\bibfnamefont
  {W.}~\bibnamefont {Wegscheider}}, \bibinfo {author} {\bibfnamefont
  {A.}~\bibnamefont {Wallraff}}, \bibinfo {author} {\bibfnamefont
  {K.}~\bibnamefont {Ensslin}}, \ and\ \bibinfo {author} {\bibfnamefont
  {T.}~\bibnamefont {Ihn}},\ }\bibfield  {title} {\enquote {\bibinfo {title}
  {Coherent spin--photon coupling using a resonant exchange qubit},}\ }\href
  {https://doi.org/10.1038/s41586-018-0365-y} {\bibfield  {journal} {\bibinfo
  {journal} {Nature}\ }\textbf {\bibinfo {volume} {560}},\ \bibinfo {pages}
  {179--184} (\bibinfo {year} {2018})}\BibitemShut {NoStop}%
\bibitem [{\citenamefont {Cerfontaine}\ \emph {et~al.}(2020)\citenamefont
  {Cerfontaine}, \citenamefont {Botzem}, \citenamefont {Ritzmann},
  \citenamefont {Humpohl}, \citenamefont {Ludwig}, \citenamefont {Schuh},
  \citenamefont {Bougeard}, \citenamefont {Wieck},\ and\ \citenamefont
  {Bluhm}}]{cerfontaine2020closed}%
  \BibitemOpen
  \bibfield  {author} {\bibinfo {author} {\bibfnamefont {P.}~\bibnamefont
  {Cerfontaine}}, \bibinfo {author} {\bibfnamefont {T.}~\bibnamefont {Botzem}},
  \bibinfo {author} {\bibfnamefont {J.}~\bibnamefont {Ritzmann}}, \bibinfo
  {author} {\bibfnamefont {S.~S.}\ \bibnamefont {Humpohl}}, \bibinfo {author}
  {\bibfnamefont {A.}~\bibnamefont {Ludwig}}, \bibinfo {author} {\bibfnamefont
  {D.}~\bibnamefont {Schuh}}, \bibinfo {author} {\bibfnamefont
  {D.}~\bibnamefont {Bougeard}}, \bibinfo {author} {\bibfnamefont {A.~D.}\
  \bibnamefont {Wieck}}, \ and\ \bibinfo {author} {\bibfnamefont
  {H.}~\bibnamefont {Bluhm}},\ }\bibfield  {title} {\enquote {\bibinfo {title}
  {Closed-loop control of a gaas-based singlet-triplet spin qubit with 99.5\%
  gate fidelity and low leakage},}\ }\href
  {https://doi.org/10.1038/s41467-020-17865-3} {\bibfield  {journal} {\bibinfo
  {journal} {Nature Communications}\ }\textbf {\bibinfo {volume} {11}},\
  \bibinfo {pages} {1--6} (\bibinfo {year} {2020})}\BibitemShut {NoStop}%
\bibitem [{\citenamefont {Kratochwil}\ \emph {et~al.}(2021)\citenamefont
  {Kratochwil}, \citenamefont {Koski}, \citenamefont {Landig}, \citenamefont
  {Scarlino}, \citenamefont {Abadillo-Uriel}, \citenamefont {Reichl},
  \citenamefont {Coppersmith}, \citenamefont {Wegscheider}, \citenamefont
  {Friesen}, \citenamefont {Wallraff}, \citenamefont {Ihn},\ and\ \citenamefont
  {Ensslin}}]{kratochwil2020realization}%
  \BibitemOpen
  \bibfield  {author} {\bibinfo {author} {\bibfnamefont {B.}~\bibnamefont
  {Kratochwil}}, \bibinfo {author} {\bibfnamefont {J.~V.}\ \bibnamefont
  {Koski}}, \bibinfo {author} {\bibfnamefont {A.~J.}\ \bibnamefont {Landig}},
  \bibinfo {author} {\bibfnamefont {P.}~\bibnamefont {Scarlino}}, \bibinfo
  {author} {\bibfnamefont {J.~C.}\ \bibnamefont {Abadillo-Uriel}}, \bibinfo
  {author} {\bibfnamefont {C.}~\bibnamefont {Reichl}}, \bibinfo {author}
  {\bibfnamefont {S.~N.}\ \bibnamefont {Coppersmith}}, \bibinfo {author}
  {\bibfnamefont {W.}~\bibnamefont {Wegscheider}}, \bibinfo {author}
  {\bibfnamefont {Mark}\ \bibnamefont {Friesen}}, \bibinfo {author}
  {\bibfnamefont {A.}~\bibnamefont {Wallraff}}, \bibinfo {author}
  {\bibfnamefont {T.}~\bibnamefont {Ihn}}, \ and\ \bibinfo {author}
  {\bibfnamefont {K.}~\bibnamefont {Ensslin}},\ }\bibfield  {title} {\enquote
  {\bibinfo {title} {Charge qubit in a triple quantum dot with tunable
  coherence},}\ }\href {\doibase 10.1103/PhysRevResearch.3.013171} {\bibfield
  {journal} {\bibinfo  {journal} {Physical Review Research}\ }\textbf {\bibinfo
  {volume} {3}},\ \bibinfo {pages} {013171} (\bibinfo {year}
  {2021})}\BibitemShut {NoStop}%
\bibitem [{\citenamefont {Recher}\ \emph {et~al.}(2009)\citenamefont {Recher},
  \citenamefont {Nilsson}, \citenamefont {Burkard},\ and\ \citenamefont
  {Trauzettel}}]{recher_2009}%
  \BibitemOpen
  \bibfield  {author} {\bibinfo {author} {\bibfnamefont {P.}~\bibnamefont
  {Recher}}, \bibinfo {author} {\bibfnamefont {J.}~\bibnamefont {Nilsson}},
  \bibinfo {author} {\bibfnamefont {G.}~\bibnamefont {Burkard}}, \ and\
  \bibinfo {author} {\bibfnamefont {B.}~\bibnamefont {Trauzettel}},\ }\bibfield
   {title} {\enquote {\bibinfo {title} {Bound states and magnetic field induced
  valley splitting in gate-tunable graphene quantum dots},}\ }\href
  {https://doi.org/10.1103/PhysRevB.79.085407} {\bibfield  {journal} {\bibinfo
  {journal} {Physical Review B}\ }\textbf {\bibinfo {volume} {79}},\ \bibinfo
  {pages} {085407} (\bibinfo {year} {2009})}\BibitemShut {NoStop}%
\bibitem [{\citenamefont {McCann}(2006)}]{mccan2006}%
  \BibitemOpen
  \bibfield  {author} {\bibinfo {author} {\bibfnamefont {E.}~\bibnamefont
  {McCann}},\ }\bibfield  {title} {\enquote {\bibinfo {title} {Asymmetry gap in
  the electronic band structure of bilayer graphene},}\ }\href {\doibase
  10.1103/PhysRevB.74.161403} {\bibfield  {journal} {\bibinfo  {journal}
  {Physical Review B}\ }\textbf {\bibinfo {volume} {74}},\ \bibinfo {pages}
  {161403} (\bibinfo {year} {2006})}\BibitemShut {NoStop}%
\bibitem [{\citenamefont {Zarenia}\ \emph {et~al.}(2013)\citenamefont
  {Zarenia}, \citenamefont {Partoens}, \citenamefont {Chakraborty},\ and\
  \citenamefont {Peeters}}]{zarenia2013}%
  \BibitemOpen
  \bibfield  {author} {\bibinfo {author} {\bibfnamefont {M.}~\bibnamefont
  {Zarenia}}, \bibinfo {author} {\bibfnamefont {B.}~\bibnamefont {Partoens}},
  \bibinfo {author} {\bibfnamefont {T.}~\bibnamefont {Chakraborty}}, \ and\
  \bibinfo {author} {\bibfnamefont {F.~M.}\ \bibnamefont {Peeters}},\
  }\bibfield  {title} {\enquote {\bibinfo {title} {Electron-electron
  interactions in bilayer graphene quantum dots},}\ }\href {\doibase
  10.1103/PhysRevB.88.245432} {\bibfield  {journal} {\bibinfo  {journal}
  {Physical Review B}\ }\textbf {\bibinfo {volume} {88}},\ \bibinfo {pages}
  {245432} (\bibinfo {year} {2013})}\BibitemShut {NoStop}%
\bibitem [{\citenamefont {Rozhkov}\ \emph {et~al.}(2016)\citenamefont
  {Rozhkov}, \citenamefont {Sboychakov}, \citenamefont {Rakhmanov},\ and\
  \citenamefont {Nori}}]{rozhkov_2016}%
  \BibitemOpen
  \bibfield  {author} {\bibinfo {author} {\bibfnamefont {A.V.}\ \bibnamefont
  {Rozhkov}}, \bibinfo {author} {\bibfnamefont {A.O.}\ \bibnamefont
  {Sboychakov}}, \bibinfo {author} {\bibfnamefont {A.L.}\ \bibnamefont
  {Rakhmanov}}, \ and\ \bibinfo {author} {\bibfnamefont {Franco}\ \bibnamefont
  {Nori}},\ }\bibfield  {title} {\enquote {\bibinfo {title} {Electronic
  properties of graphene-based bilayer systems},}\ }\href {\doibase
  10.1016/j.physrep.2016.07.003} {\bibfield  {journal} {\bibinfo  {journal}
  {Physics Reports}\ }\textbf {\bibinfo {volume} {648}},\ \bibinfo {pages}
  {1--104} (\bibinfo {year} {2016})}\BibitemShut {NoStop}%
\bibitem [{\citenamefont {van~der Maaten}\ and\ \citenamefont
  {Hinton}(2008)}]{maaten_2008_tsne_main}%
  \BibitemOpen
  \bibfield  {author} {\bibinfo {author} {\bibfnamefont {L.}~\bibnamefont
  {van~der Maaten}}\ and\ \bibinfo {author} {\bibfnamefont {G.}~\bibnamefont
  {Hinton}},\ }\bibfield  {title} {\enquote {\bibinfo {title} {Visualizing data
  using t-sne},}\ }\href {http://jmlr.org/papers/v9/vandermaaten08a.html}
  {\bibfield  {journal} {\bibinfo  {journal} {Journal of Machine Learning
  Research}\ }\textbf {\bibinfo {volume} {9}},\ \bibinfo {pages} {2579--2605}
  (\bibinfo {year} {2008})}\BibitemShut {NoStop}%
\bibitem [{\citenamefont {Li}\ \emph {et~al.}(2013)\citenamefont {Li},
  \citenamefont {Tang}, \citenamefont {Omidvar}, \citenamefont {Yang},\ and\
  \citenamefont {Qin}}]{LSGO_benchmark_2013}%
  \BibitemOpen
  \bibfield  {author} {\bibinfo {author} {\bibfnamefont {X.}~\bibnamefont
  {Li}}, \bibinfo {author} {\bibfnamefont {K.}~\bibnamefont {Tang}}, \bibinfo
  {author} {\bibfnamefont {M.~N.}\ \bibnamefont {Omidvar}}, \bibinfo {author}
  {\bibfnamefont {Z.}~\bibnamefont {Yang}}, \ and\ \bibinfo {author}
  {\bibfnamefont {K.}~\bibnamefont {Qin}},\ }\bibfield  {title} {\enquote
  {\bibinfo {title} {{Benchmark Functions for the CEC'2013 Special Session and
  Competition on Large-Scale Global Optimization}},}\ }\href
  {https://www.researchgate.net/publication/261562928_Benchmark_Functions_for_the_CEC'2013_Special_Session_and_Competition_on_Large-Scale_Global_Optimization}
  {\  (\bibinfo {year} {2013})}\BibitemShut {NoStop}%
\bibitem [{\citenamefont {Cao}\ \emph {et~al.}(2003)\citenamefont {Cao},
  \citenamefont {Li}, \citenamefont {Petzold},\ and\ \citenamefont
  {Serban}}]{cao_2003_adjoint_method}%
  \BibitemOpen
  \bibfield  {author} {\bibinfo {author} {\bibfnamefont {Y.}~\bibnamefont
  {Cao}}, \bibinfo {author} {\bibfnamefont {S.}~\bibnamefont {Li}}, \bibinfo
  {author} {\bibfnamefont {L.}~\bibnamefont {Petzold}}, \ and\ \bibinfo
  {author} {\bibfnamefont {R.}~\bibnamefont {Serban}},\ }\bibfield  {title}
  {\enquote {\bibinfo {title} {Adjoint sensitivity analysis for
  differential-algebraic equations: The adjoint {DAE} system and its numerical
  solution},}\ }\href {\doibase 10.1137/S1064827501380630} {\bibfield
  {journal} {\bibinfo  {journal} {SIAM Journal on Scientific Computing}\
  }\textbf {\bibinfo {volume} {24}},\ \bibinfo {pages} {1076--1089} (\bibinfo
  {year} {2003})}\BibitemShut {NoStop}%
\bibitem [{\citenamefont {Feynman}(1939)}]{feynman_1939_HF}%
  \BibitemOpen
  \bibfield  {author} {\bibinfo {author} {\bibfnamefont {R.~P.}\ \bibnamefont
  {Feynman}},\ }\bibfield  {title} {\enquote {\bibinfo {title} {Forces in
  molecules},}\ }\href {\doibase 10.1103/PhysRev.56.340} {\bibfield  {journal}
  {\bibinfo  {journal} {Physical Review}\ }\textbf {\bibinfo {volume} {56}},\
  \bibinfo {pages} {340--343} (\bibinfo {year} {1939})}\BibitemShut {NoStop}%
\bibitem [{\citenamefont {Arnoud}\ \emph {et~al.}(2019)\citenamefont {Arnoud},
  \citenamefont {Guvenen},\ and\ \citenamefont
  {Kleineberg}}]{Arnoud2019BenchmarkingGO}%
  \BibitemOpen
  \bibfield  {author} {\bibinfo {author} {\bibfnamefont {A.}~\bibnamefont
  {Arnoud}}, \bibinfo {author} {\bibfnamefont {F.}~\bibnamefont {Guvenen}}, \
  and\ \bibinfo {author} {\bibfnamefont {T.}~\bibnamefont {Kleineberg}},\
  }\bibfield  {title} {\enquote {\bibinfo {title} {Benchmarking {G}lobal
  {O}ptimizers},}\ }\href {\doibase 10.3386/w26340} {\bibfield  {journal}
  {\bibinfo  {journal} {Econometrics: Econometric \& Statistical Methods -
  Special Topics eJournal}\ } (\bibinfo {year} {2019}),\
  10.3386/w26340}\BibitemShut {NoStop}%
\bibitem [{\citenamefont {Kaelo}\ and\ \citenamefont
  {Ali}(2006)}]{kaelo_ali_2006_crs-lm}%
  \BibitemOpen
  \bibfield  {author} {\bibinfo {author} {\bibfnamefont {P.}~\bibnamefont
  {Kaelo}}\ and\ \bibinfo {author} {\bibfnamefont {M.}~\bibnamefont {Ali}},\
  }\bibfield  {title} {\enquote {\bibinfo {title} {{Some Variants of the
  Controlled Random Search Algorithm for Global Optimization}},}\ }\href
  {https://dl.acm.org/doi/10.1007/s10957-006-9101-0} {\bibfield  {journal}
  {\bibinfo  {journal} {Journal of Optimization Theory and Applications}\
  }\textbf {\bibinfo {volume} {130}},\ \bibinfo {pages} {253--264} (\bibinfo
  {year} {2006})}\BibitemShut {NoStop}%
\bibitem [{\citenamefont {Johnson}()}]{johnson_nlopt}%
  \BibitemOpen
  \bibfield  {author} {\bibinfo {author} {\bibfnamefont {S.~G.}\ \bibnamefont
  {Johnson}},\ }\href@noop {} {\enquote {\bibinfo {title} {{NLopt} - a library
  for nonlinear local and global optimization},}\ }\bibinfo {howpublished}
  {\url{http://github.com/stevengj/nlopt}}\BibitemShut {NoStop}%
\bibitem [{\citenamefont {Bucko}\ \emph {et~al.}()\citenamefont {Bucko} \emph
  {et~al.}}]{BGQD_code}%
  \BibitemOpen
  \bibfield  {author} {\bibinfo {author} {\bibfnamefont {J.}~\bibnamefont
  {Bucko}} \emph {et~al.},\ }\href@noop {} {\enquote {\bibinfo {title}
  {Automated reconstruction of bound states in bilayer graphene quantum
  dots},}\ }\bibinfo {howpublished}
  {\url{https://gitlab.com/QMAI/papers/bilayergrapheneqds}}\BibitemShut
  {NoStop}%
\bibitem [{\citenamefont {Knothe}\ and\ \citenamefont
  {Fal'ko}(2020)}]{knothe_2020}%
  \BibitemOpen
  \bibfield  {author} {\bibinfo {author} {\bibfnamefont {Angelika}\
  \bibnamefont {Knothe}}\ and\ \bibinfo {author} {\bibfnamefont {Vladimir}\
  \bibnamefont {Fal'ko}},\ }\bibfield  {title} {\enquote {\bibinfo {title}
  {Quartet states in two-electron quantum dots in bilayer graphene},}\ }\href
  {\doibase 10.1103/PhysRevB.101.235423} {\bibfield  {journal} {\bibinfo
  {journal} {Phys. Rev. B}\ }\textbf {\bibinfo {volume} {101}},\ \bibinfo
  {pages} {235423} (\bibinfo {year} {2020})}\BibitemShut {NoStop}%
\bibitem [{\citenamefont {Garreis}\ \emph {et~al.}(2021)\citenamefont
  {Garreis}, \citenamefont {Knothe}, \citenamefont {Tong}, \citenamefont
  {Eich}, \citenamefont {Gold}, \citenamefont {Watanabe}, \citenamefont
  {Taniguchi}, \citenamefont {Fal'ko}, \citenamefont {Ihn}, \citenamefont
  {Ensslin},\ and\ \citenamefont {Kurzmann}}]{garreis_2021}%
  \BibitemOpen
  \bibfield  {author} {\bibinfo {author} {\bibfnamefont {R.}~\bibnamefont
  {Garreis}}, \bibinfo {author} {\bibfnamefont {A.}~\bibnamefont {Knothe}},
  \bibinfo {author} {\bibfnamefont {C.}~\bibnamefont {Tong}}, \bibinfo {author}
  {\bibfnamefont {M.}~\bibnamefont {Eich}}, \bibinfo {author} {\bibfnamefont
  {C.}~\bibnamefont {Gold}}, \bibinfo {author} {\bibfnamefont {K.}~\bibnamefont
  {Watanabe}}, \bibinfo {author} {\bibfnamefont {T.}~\bibnamefont {Taniguchi}},
  \bibinfo {author} {\bibfnamefont {V.}~\bibnamefont {Fal'ko}}, \bibinfo
  {author} {\bibfnamefont {T.}~\bibnamefont {Ihn}}, \bibinfo {author}
  {\bibfnamefont {K.}~\bibnamefont {Ensslin}}, \ and\ \bibinfo {author}
  {\bibfnamefont {A.}~\bibnamefont {Kurzmann}},\ }\bibfield  {title} {\enquote
  {\bibinfo {title} {Shell filling and trigonal warping in graphene quantum
  dots},}\ }\href {\doibase 10.1103/PhysRevLett.126.147703} {\bibfield
  {journal} {\bibinfo  {journal} {Phys. Rev. Lett.}\ }\textbf {\bibinfo
  {volume} {126}},\ \bibinfo {pages} {147703} (\bibinfo {year}
  {2021})}\BibitemShut {NoStop}%
\bibitem [{\citenamefont {Möller}\ \emph {et~al.}(2021)\citenamefont
  {Möller}, \citenamefont {Banszerus}, \citenamefont {Knothe}, \citenamefont
  {Steiner}, \citenamefont {Icking}, \citenamefont {Trellenkamp}, \citenamefont
  {Lentz}, \citenamefont {Watanabe}, \citenamefont {Taniguchi}, \citenamefont
  {Glazman}, \citenamefont {Fal'ko}, \citenamefont {Volk},\ and\ \citenamefont
  {Stampfer}}]{moller_2021}%
  \BibitemOpen
  \bibfield  {author} {\bibinfo {author} {\bibfnamefont {S.}~\bibnamefont
  {Möller}}, \bibinfo {author} {\bibfnamefont {L.}~\bibnamefont {Banszerus}},
  \bibinfo {author} {\bibfnamefont {A.}~\bibnamefont {Knothe}}, \bibinfo
  {author} {\bibfnamefont {C.}~\bibnamefont {Steiner}}, \bibinfo {author}
  {\bibfnamefont {E.}~\bibnamefont {Icking}}, \bibinfo {author} {\bibfnamefont
  {S.}~\bibnamefont {Trellenkamp}}, \bibinfo {author} {\bibfnamefont
  {F.}~\bibnamefont {Lentz}}, \bibinfo {author} {\bibfnamefont
  {K.}~\bibnamefont {Watanabe}}, \bibinfo {author} {\bibfnamefont
  {T.}~\bibnamefont {Taniguchi}}, \bibinfo {author} {\bibfnamefont
  {L.{\hspace{0.167em}}I.}\ \bibnamefont {Glazman}}, \bibinfo {author}
  {\bibfnamefont {V.{\hspace{0.167em}}I.}\ \bibnamefont {Fal'ko}}, \bibinfo
  {author} {\bibfnamefont {C.}~\bibnamefont {Volk}}, \ and\ \bibinfo {author}
  {\bibfnamefont {C.}~\bibnamefont {Stampfer}},\ }\bibfield  {title} {\enquote
  {\bibinfo {title} {Probing two-electron multiplets in bilayer graphene
  quantum dots},}\ }\href {\doibase 10.1103/physrevlett.127.256802} {\bibfield
  {journal} {\bibinfo  {journal} {Physical Review Letters}\ }\textbf {\bibinfo
  {volume} {127}} (\bibinfo {year} {2021}),\
  10.1103/physrevlett.127.256802}\BibitemShut {NoStop}%
\bibitem [{\citenamefont {Knothe}\ \emph {et~al.}(2022)\citenamefont {Knothe},
  \citenamefont {Glazman},\ and\ \citenamefont {Fal'ko}}]{knothe_2022}%
  \BibitemOpen
  \bibfield  {author} {\bibinfo {author} {\bibfnamefont {Angelika}\
  \bibnamefont {Knothe}}, \bibinfo {author} {\bibfnamefont {Leonid~I}\
  \bibnamefont {Glazman}}, \ and\ \bibinfo {author} {\bibfnamefont
  {Vladimir~I}\ \bibnamefont {Fal'ko}},\ }\bibfield  {title} {\enquote
  {\bibinfo {title} {Tunneling theory for a bilayer graphene quantum dot's
  single- and two-electron states},}\ }\href {\doibase
  10.1088/1367-2630/ac5d00} {\bibfield  {journal} {\bibinfo  {journal} {New
  Journal of Physics}\ }\textbf {\bibinfo {volume} {24}},\ \bibinfo {pages}
  {043003} (\bibinfo {year} {2022})}\BibitemShut {NoStop}%
\bibitem [{\citenamefont {Savitzky}\ and\ \citenamefont
  {Golay}(1964)}]{sav_gol_1964}%
  \BibitemOpen
  \bibfield  {author} {\bibinfo {author} {\bibfnamefont {Abraham.}\
  \bibnamefont {Savitzky}}\ and\ \bibinfo {author} {\bibfnamefont {M.~J.~E.}\
  \bibnamefont {Golay}},\ }\bibfield  {title} {\enquote {\bibinfo {title}
  {Smoothing and differentiation of data by simplified least squares
  procedures.}}\ }\href {\doibase 10.1021/ac60214a047} {\bibfield  {journal}
  {\bibinfo  {journal} {Analytical Chemistry}\ }\textbf {\bibinfo {volume}
  {36}},\ \bibinfo {pages} {1627--1639} (\bibinfo {year} {1964})},\ \Eprint
  {http://arxiv.org/abs/https://doi.org/10.1021/ac60214a047}
  {https://doi.org/10.1021/ac60214a047} \BibitemShut {NoStop}%
\bibitem [{sob()}]{sobol_seq}%
  \BibitemOpen
  \href@noop {} {\enquote {\bibinfo {title} {Sobol sequence implementation in
  python},}\ }\bibinfo {howpublished}
  {\url{https://github.com/naught101/sobol_seq}}\BibitemShut {NoStop}%
\bibitem [{\citenamefont {Falkner}\ \emph {et~al.}(2018)\citenamefont
  {Falkner}, \citenamefont {Klein},\ and\ \citenamefont
  {Hutter}}]{falkner2018bohb}%
  \BibitemOpen
  \bibfield  {author} {\bibinfo {author} {\bibfnamefont {S.}~\bibnamefont
  {Falkner}}, \bibinfo {author} {\bibfnamefont {A.}~\bibnamefont {Klein}}, \
  and\ \bibinfo {author} {\bibfnamefont {F.}~\bibnamefont {Hutter}},\
  }\bibfield  {title} {\enquote {\bibinfo {title} {{BOHB}: Robust and efficient
  hyperparameter optimization at scale},}\ }in\ \href
  {http://proceedings.mlr.press/v80/falkner18a.html} {\emph {\bibinfo
  {booktitle} {Proceedings of the 35th International Conference on Machine
  Learning}}},\ \bibinfo {series} {Proceedings of Machine Learning Research},
  Vol.~\bibinfo {volume} {80},\ \bibinfo {editor} {edited by\ \bibinfo {editor}
  {\bibfnamefont {J.}~\bibnamefont {Dy}}\ and\ \bibinfo {editor} {\bibfnamefont
  {A.}~\bibnamefont {Krause}}}\ (\bibinfo  {publisher} {PMLR},\ \bibinfo
  {address} {Stockholmsmässan, Stockholm Sweden},\ \bibinfo {year} {2018})\
  pp.\ \bibinfo {pages} {1437--1446}\BibitemShut {NoStop}%
\end{thebibliography}%

\end{document}